\documentclass[10pt,journal,twocolumns]{IEEEtran}
\usepackage{subfigure}
\usepackage{amssymb}
\usepackage{amsthm}
\usepackage{amsmath}
\usepackage{hyperref}
\usepackage{graphicx}
\usepackage{colortbl,dcolumn}
\usepackage{booktabs}
\usepackage{xcolor}
\usepackage{cite}
\usepackage{threeparttable}

\newtheorem{definition}{Definition}
\newtheorem{lemma}{Lemma}
\newtheorem{theorem}{Theorem}
\newtheorem{corollary}{Corollary}
\newtheorem{remark}{Remark}

\newcommand{\lemref}[1]{Lemma~\ref{#1}}
\newcommand{\thmref}[1]{Theorem~\ref{#1}}

\newcommand{\secref}[1]{Section~\ref{#1}}

\usepackage{mathtools} 

\DeclarePairedDelimiter{\brkt}{[}{]}
\newcommand{\Expect}{\operatorname{\mathbb{E}}\brkt}
\newcommand{\Stat}{\operatorname{\mathbb{S}}\brkt}
\newcommand{\Cov}{\operatorname{Cov}\brkt}
\DeclarePairedDelimiter{\rbrkt}{(}{)}
\newcommand{\Covv}{\operatorname{Cov}\rbrkt}

\ifCLASSOPTIONcompsoc
\else
\fi
\ifCLASSINFOpdf
\else
\fi

\graphicspath{{pdfs/},{eps/},{authorphotos/}}

\begin{document}
%
\title{Interference Prediction in Mobile Ad Hoc Networks with a General Mobility Model}
%
%
%
%

\author{Yirui~Cong,~\IEEEmembership{Student~Member,~IEEE,}
        Xiangyun~Zhou,~\IEEEmembership{Member,~IEEE,}
        and~Rodney~A.~Kennedy,~\IEEEmembership{Fellow,~IEEE}
\IEEEcompsocitemizethanks{\IEEEcompsocthanksitem
Y. Cong, X. Zhou, and R. A. Kennedy are with the Research School
of Engineering, Australian National University, Australia (Email: \{yirui.cong, xiangyun.zhou, Rodney.Kennedy\}@anu.edu.au).\protect
}
}

%
%

\markboth{}%
{Cong \MakeLowercase{\textit{et al.}}: Interference Prediction in Mobile Ad Hoc Networks with a General Mobility Model}
%



\IEEEtitleabstractindextext{%
\begin{abstract}
In a mobile ad hoc network (MANET), effective prediction of time-varying interferences can enable adaptive transmission designs and therefore improve the communication performance.
This paper investigates interference prediction in MANETs with a finite number of nodes by proposing and using a general-order linear model for node mobility.
The proposed mobility model can well approximate node dynamics of practical MANETs.
%
%
In contrast to previous studies on interference statistics, we are able through this model to give a best estimate of the time-varying interference at any time rather than long-term average effects.
Specifically, we propose a compound Gaussian point process functional as a general framework to obtain analytical results on the mean value and moment-generating function of the interference prediction.
With a series form of this functional, we give the necessary and sufficient condition for when the prediction is essentially equivalent to that from a Binomial Point Process (BPP) network in the limit as time goes to infinity.
These conditions permit one to rigorously determine when the commonly used BPP approximations are valid. Finally, our simulation results corroborate the effectiveness and accuracy of the analytical results on interference prediction and also show the advantages of our method in dealing with complex mobilities.
\end{abstract}

\begin{IEEEkeywords}
Interference Prediction, General-order Mobility, Compound Gaussian Point Process Functional, Gaussian BPP, Mobile Ad hoc Networks.
\end{IEEEkeywords}}

\maketitle

\IEEEdisplaynontitleabstractindextext

%
\IEEEpeerreviewmaketitle

\section{Introduction}

\subsection{Motivation and Related Work}

\IEEEPARstart{I}{n} MANETs interference plays a pivotal role, through the Signal to Interference Ratio (SIR), in contributing to the Quality of Service (QoS).
In contrast to static wireless networks, the distances between interferers and receiver change dynamically during the times of communications because of the mobilities of interferers and receiver.
As a result, the received signal is affected by fluctuating interferences generated by these mobile nodes.

Interference analysis can help to discover and exploit the regularity of these time-vary interferences.
For example, it helps to understand the statistical performance of communication under such interference, e.g., outage probability.
Due to disturbances on nodes' movement or our incomplete information of the node locations, the trajectories of these nodes are often accompanied with uncertainties when analyzing interference in MANETs.
Stochastic Geometry~\cite{HaenggiM2012_BOOK,BaccelliF2009_BOOK_part1} is a powerful tool for describing the random pattern of mobile nodes as a Point Process (PP) at each time instant.
For mobile networks with high mobility nodes, the authors in~\cite{HaenggiM2010} analyzed the local delay, which is a functional of the interference Moment Generating Function (MGF), by modeling node locations as a Poisson Point Process (PPP).
Indeed, if one just focuses on one time instant, the existing results on interference analysis in static networks can be directly used in highly mobile networks (e.g., for PPP networks~\cite{HaenggiM2010,HaenggiM2009_FTN,HaenggiM2009,TanbourgiR2014} or for Binomial Point Process (BPP) networks~\cite{SrinivasaS2010,GuoJ2014}).
Highly mobile networks imply that the node locations at two different time instants have almost no correlation, if the node velocities are sufficiently large.
Nevertheless, for most practical cases, we are more interested in interference analysis in MANETs with finite nodal velocities.

In order to analyze interference in MANETs with finite node velocities, mobility models are required to capture the node locations at each time instant.
A summary of mobility models and their corresponding point processes in prior studies on interference statistics of MANETs is provided in Table~\ref{tab:Summary of Mobility Models and Their Corresponding Point Processes in Prior Studies on Interference Statistics of MANETs}.
\begin{table*}[htb]
\centering
\begin{threeparttable}
\caption{Summary of Mobility Models and Their Corresponding Point Processes in Prior Studies on Interference Statistics of MANETs\label{tab:Summary of Mobility Models and Their Corresponding Point Processes in Prior Studies on Interference Statistics of MANETs}}
\centering
\begin{small}
\begin{tabular}{lp{4cm}p{8cm}}
\toprule
Reference & Mobility Model & Point Process\\
\midrule
\cite{HaenggiM2010,HaenggiM2009_FTN,HaenggiM2009,TanbourgiR2014,GantiR2009,SchilcherU2012,HaenggiM2012CL,CrismaniA2015,SchilcherU2013}& Highly mobility networks & PPP (time independent)\\
\cite{GulatiK2012}& Highly mobility networks with randomly actived interferes & PPP (time homogenous for sufficient large time)\\
\cite{SrinivasaS2010,GuoJ2014}& Highly mobility networks & BPP (time independent)\\
\cite{GongZ2013,GongZ2011,GongZ2014}& Constrained i.i.d. mobility & PPP (time homogenous, non-Markov~\cite{GongZ2014})\\
\cite{GongZ2010,GongZ2011,GongZ2014}& Random walk & PPP (time homogenous)\\
\cite{GongZ2011,GongZ2014}& Brownian motion & PPP (time homogenous)\\
\cite{GongZ2010,GongZ2011,GongZ2014}& Random waypoint & PPP (time homogenous) with quadratic polynomial intensity\\
\bottomrule
\end{tabular}
\begin{tablenotes}\footnotesize
\item Time independent: the locations of nodes change independently from one time instant to the next.
\item Time homogenous: the locations of nodes are correlated between different time instants but have the same pdf.
\end{tablenotes}
\end{small}
\end{threeparttable}
\end{table*}
By employing random walk, Brownian motion and random waypoint models\footnote{The constrained i.i.d.~mobility model in~\cite{GongZ2013} is similar to the highly mobile network, thus we mainly discuss the interference analysis in MANETs under random walk, Brownian motion and random waypoint.}, the statistics of interference were analyzed in~\cite{GongZ2010,GongZ2011,GongZ2014}.
For random walk and Brownian motion models, the approximate distribution of aggregate interference was given in~\cite{GongZ2010,GongZ2014}.
The mean of the aggregate interference was analyzed in~\cite{GongZ2014}, and upper bounds in time-correlation for aggregate interference and outage probability were given in~\cite{GongZ2011} and~\cite{GongZ2014}.
For the random waypoint model, similarly, the approximated distribution of aggregate interference was given in~\cite{GongZ2010,GongZ2014}, and the mean of the aggregate interference was analyzed in~\cite{GongZ2014}.


However, the existing analysis methods only considered special or limiting scenarios where statistics of interference are identical at every time instant.
For the random walk and Brownian motion models in~\cite{GongZ2010,GongZ2014}, the means or pdfs (probability density functions) of aggregate interference at every time instant are the same due to the assumption that the initial node distribution is uniform.
Consequently, node dynamics do not provide any useful information in the interference analysis.
For the random waypoint model in~\cite{GongZ2010,GongZ2014}, the idea was to wait an infinite time for the node distribution to converge to one limiting distribution that can be regarded as a PPP with a quadratic polynomial intensity~\cite{BettstetterC2003}.
Based on that kind of PPP, mean, outage and approximated distribution for aggregate interference were given.
As a result, not only do node dynamics not contribute to the analysis, but also the results are only valid in infinite time.
It should be noted that in many applications what we would like to know are the statistics of aggregate interference within a finite time window rather than an infinite one.

Furthermore, the existing analysis on interference between multiple time instants typically focused on the time correlation~\cite{GongZ2011,GongZ2014,GantiR2009,SchilcherU2012,HaenggiM2012CL}, which provides no further information beyond a linear relation.
To design a communication strategy (e.g., transmission power control), we are interested in knowing and exploiting the statistics of interference at a future time of interest, which cannot be derived from a simple time correlation.
We assert that interference prediction can provide us more effective information than time correlation.
Beyond the simple temporal correlation,~\cite{GulatiK2012} proposed a joint temporal characteristic function of interference for multiple time instants, and~\cite{GantiR2009,CrismaniA2015,SchilcherU2013} investigated the conditional/joint outage/success probabilities over time.
Despite the comprehensive characterizations of the temporal statistics, the studies in~\cite{GantiR2009,CrismaniA2015,SchilcherU2013} assumed networks with either fixed or independent node locations from one time slot to the next.
Hence, their results cannot be used for interference prediction in realistic network with practical mobility models.

In fact, another limitation of current studies on interference in MANETs is the mobility model.
For example, the random walk, Brownian motion and random waypoint models are often inadequate to describe many kinds of mobilities in the real world, like mobilities constrained by the physical laws of acceleration and velocity~\cite{BaiF2004}.
%
%
Modelling should include all kinds of communicating objects that have the ability to move.
With the development of automation, the need for communication between robots is increasing.
%
%
For civil use, unmanned aircrafts offer new ways for commercial enterprises and public operators to increase operational efficiency, decrease costs, and enhance safety~\cite{ICUAS2013}.
The node mobilities in such scenarios have complex dynamics and cannot be captured by random walk, Brownian motion, random waypoint mobility models or even the Gauss-Markov mobility model~\cite{BaiF2004,CampT2002} that takes acceleration and velocity into account.
Therefore, it is desirable to develop a more general mobility model that is capable to describe mobile nodes governed by complex mobility dynamics, and then use it for interference prediction.

\subsection{Our Contributions}

In this work, we focus on interference prediction in MANETs with general mobilities having a finite number of nodes.
Compared to time correlation, interference prediction can give more effective information, i.e., providing the best estimate of the interference level at a future time instant based on the knowledge at the current time.
By developing a general mobility model, the predictions can be used in a wide range of MANETs.

The main contributions of this work are:

\begin{itemize}
\item We propose a General-order Linear Continuous-time (GLC) mobility model to describe the dynamics of moving nodes in practical applications.
The random walk, Brownian motion and Gauss-Markov mobility models in \cite{GongZ2010,GongZ2014,BaiF2004} can be regarded as special cases of the GLC mobility model with discretizations.
In this framework, the random walk and Brownian motion turn out to be first-order linear mobility models, and Gauss-Markov model is a second-order linear mobility model.

\item Based on the GLC mobility model, the mean and Moment-Generating Function (MGF) of interference prediction on a mobile reference point at any finite time into the future are derived and analyzed.
In order to simplify the expression for mean and MGF under different path loss functions and multipath fading, a Compound Gaussian Point Process Functional (CGPPF) is defined and expressed in a series form.

\item Apart from interference prediction at finite time into the future, we also give the necessary and sufficient condition for when our predictions converge to those from a Gaussian BPP as time goes to infinity.
This result provides a guideline on when the previous studies on interference statistics with BPP are relevant.
\end{itemize}

\subsection{Paper Organization}

In \secref{sec:Uniform Circular Motions}, we present the case of Uniform Circular Motion (UCM) as an example to motivate our general model.
In \secref{sec:GLCmod}, we define and analyze the GLC mobility model.
Additionally, the dynamic reference point is defined to study the relative node locations of interferers relative to the mobile receiver.
In \secref{sec:InterferencePredictions}, the mean and MGF of the interference prediction are analyzed.
%
The numerical examples are given in \secref{sec:SimulationExamples} to illustrate the effectiveness of our approach and corroborate our analytical results.

\subsection{Notation}

$\Expect{\cdot}$ denotes the mean of a random variable or a random vector, and $\mathbb{E}_a[\cdot]$ denotes the expectation operator with respect to $a$.
$\Cov{\cdot}$ denotes the covariance of a random vector, and
$\dot{x}(t) = {dx(t)}/{dt}$ denotes the derivative of stochastic process $x(t)$ at time $t$ in the mean square sense~\cite{OksendalB2000}.

\section{Interferences in MANETs with Uniform Circular Motions}\label{sec:Uniform Circular Motions}

In this section, we investigate the interference that arises in nodes exhibiting Uniform Circular Motion (UCM) to model UAVs carrying out a scanning task.
This simple model will be seen to be a special case of the general GLC mobility model that we develop in \secref{sec:GLCmod}.

%
%

\begin{figure}
\centering
\subfigure[]{\includegraphics [width=0.72\columnwidth, height=0.61\columnwidth]{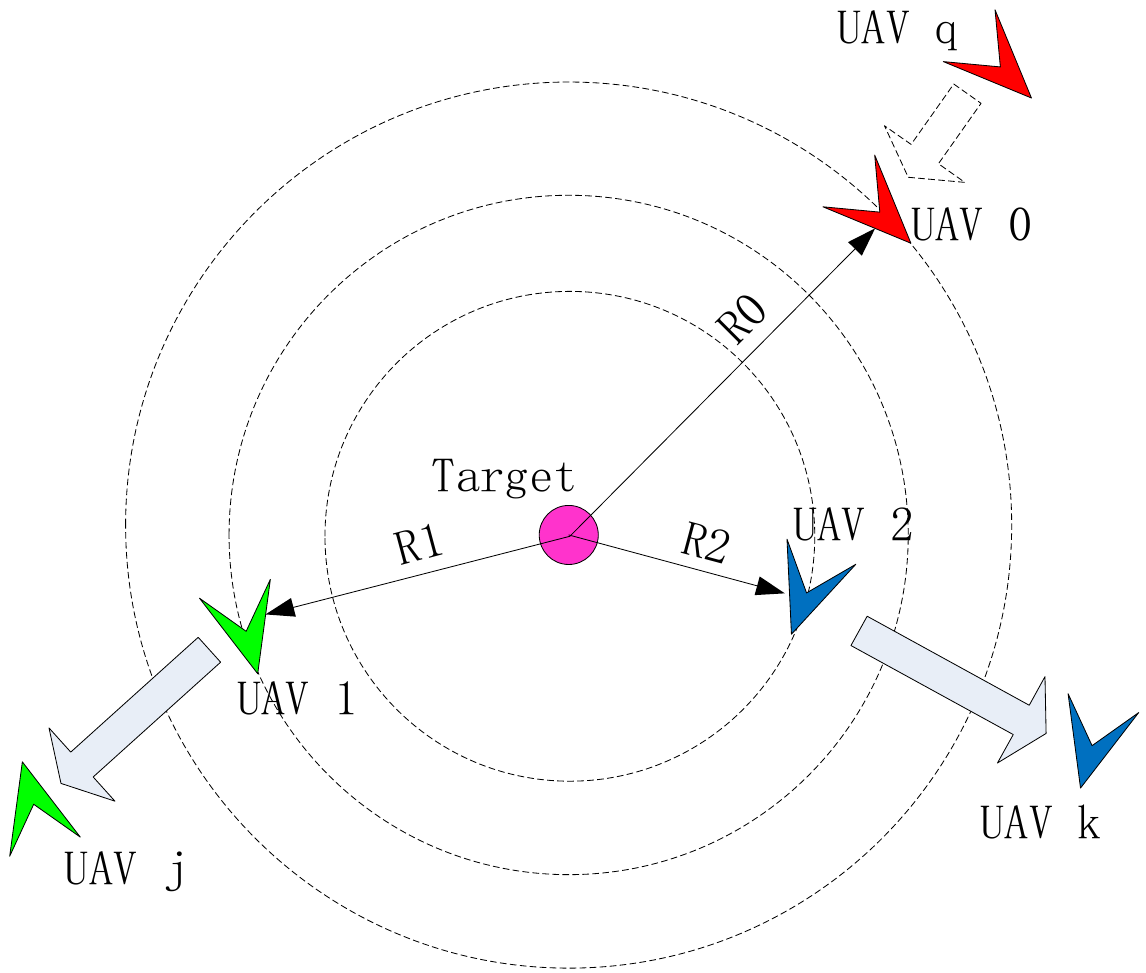}\label{fig:UAVsScan}}\\
\subfigure[]{\includegraphics [width=0.72\columnwidth, height=0.61\columnwidth]{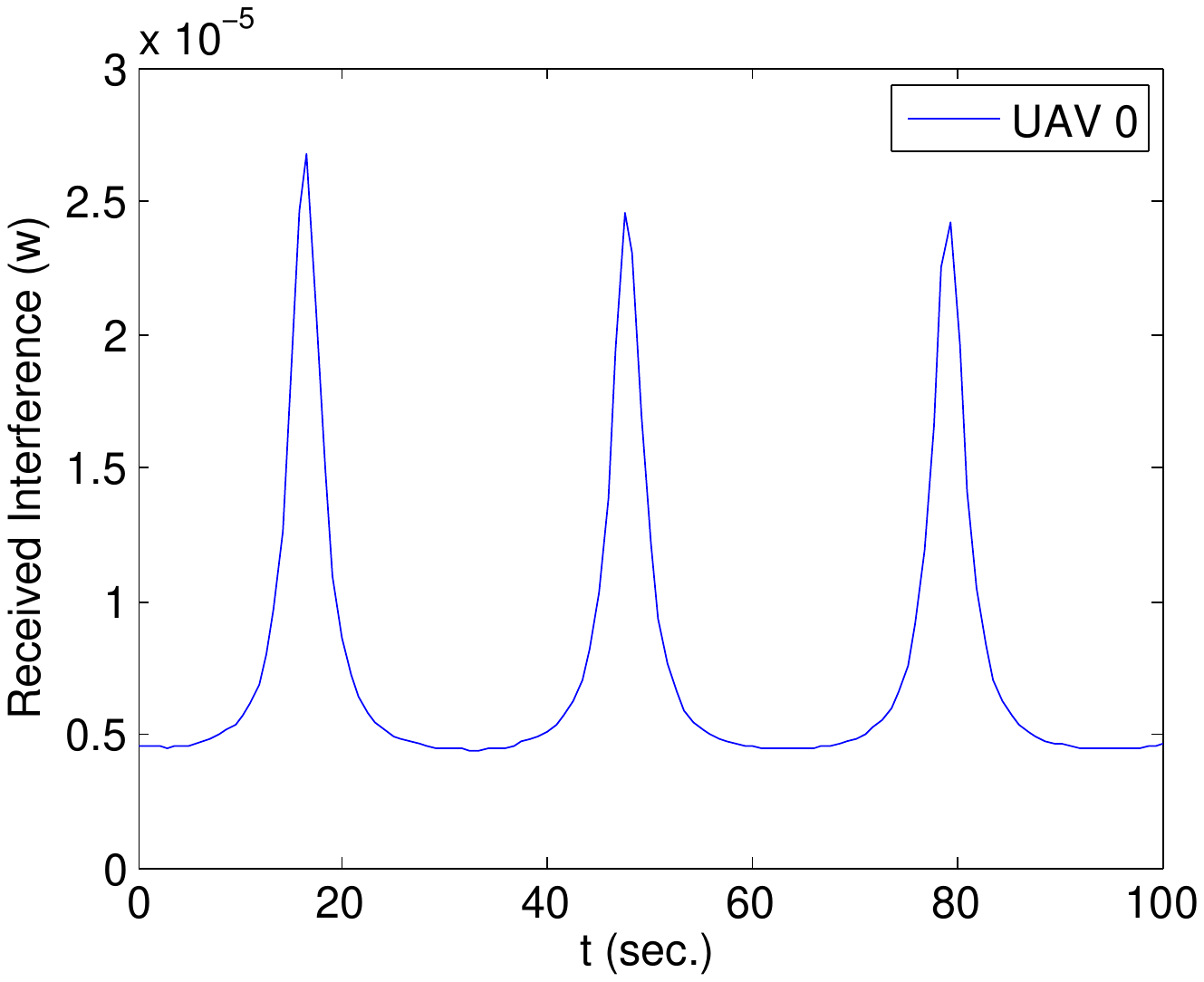}\label{fig:UAVsScanInterference}}
\caption{(a) Three UAVs scan a target by using vision sensors and share information through a wireless network. (b) The received interference UAV $0$ versus time $t$.
}
\end{figure}

Consider $N\in\mathbb{Z}^{+}$ UAVs carrying out a scanning task for a target, see Fig.~\ref{fig:UAVsScan}, which has many applications such as rescue operations, monitoring applications, etc.
This target (the central pink circle) could be a point or an area.
%
%
Due to limitations of their visual fields, each UAV can only acquire partial information about the target.
Therefore, in order to have a better understanding of the target, they need to share information using wireless communication.

Even though the dynamic model of the UAVs when scanning a target can be highly nonlinear and possibly complex, the mobility model can be approximated by UCM in the 2-D plane as two coupled differential equations
\begin{equation}
\label{eqn:UCMwithPerturbations}
\left\{
\begin{split}
{\dot{x}}_i^{(1)}(t) &= \omega_i x_i^{(2)}(t) + w_i^{(1)}(t) \\
{\dot{x}}_i^{(2)}(t) &= -\omega_i x_i^{(1)}(t) + w_i^{(2)}(t)
\end{split}
\right.
\end{equation}
where the subscript $i$ denotes the $i$th node, and $i=1,2,\dotsc,N$, where $N$ is the number of nodes in the MANET.
$(x_i^{(1)}(t),x_i^{(2)}(t))$ is the location of node $i$ in the 2D plane at time $t$, and $\omega_i$ is the UCM angular velocity, and the initial vector ${\bf x}_i(t_0) = [x_i^{(1)}(t_0),x_i^{(2)}(t_0)]^T$ determines the initial location and radius of node $i$.
$w_i^{(1)}(t)$ and $w_i^{(2)}(t)$ stand for additive disturbance due to airflow.

Consider a duration of time in which UAV 1 and 2 communicate with UAV $j$ and $k$, respectively.
For UAV 0, it tries to receive information from UAV $q$.
Therefore, there are two interferers that affect the signal reception at UAV 0.

The aggregate interference at UAV 0 is shown in Fig.~\ref{fig:UAVsScanInterference},\footnote{
Simulation Parameters: Signal power is $1$ for each UAV, and the mobility model follows~\eqref{eqn:UCMwithPerturbations} with $\omega_0 = \omega_2 = 0.1 \mathrm{rad/s}$ and $\omega_1 = -0.1 \mathrm{rad/s} $. The initial locations for UAV $0-2$ are $(500,500)$, $(-400,-300)$, $(400,0)$ (selecting the target center as the origin), thus according to~\eqref{eqn:UCMwithPerturbations} average radii are $R_0 = 500\sqrt{2} \mathrm{m}$, $R_1 = 500 \mathrm{m}$, and $R_2 = 400 \mathrm{m}$. The powers of $w_i^{(1)}(t)$ and $w_i^{(2)}(t)$ are assumed to be unit. We assume the Path loss function $r^{-2}$, where $r$ is the Euclid distance between transmitter (UAV) and the point whose interference is needed to be calculated.}
from which we see that the received aggregate interference at UAV 0 is periodically changing and does not converge to a constant value independent of time $t$.
Interference predictions should be adaptive to the node dynamics and therefore make use of the characteristics of mobility models.

%
%
%
%

\section{General-order Linear Mobility Model}\label{sec:GLCmod}

In this section, the General-order Linear Continuous-time (GLC) Mobility Model is proposed and the corresponding statistics of node distribution is given.
%

\subsection{General-order Linear Mobility Model and Node Distribution}

Consider a network having $N$ nodes in $d$-dimensional space (e.g., $d=2$ means nodes move in a 2D plane).
For each node $i$, where $i\in\{1,2,\dotsc,N\}$, we model it employing the state-space model with additive uncertainties given by the stochastic differential and algebraic equations~\cite{ChenC1999}
\begin{align}
	{\dot{\bf x}}_i(t)
		&= {\bf{A}}_i{\bf x}_i(t) + {\bf w}_i(t)\label{eqn:state-equation} \\
	{\bf{y}}_i(t)
		&= {\bf{C}}_i{\bf x}_i(t),\label{eqn:output-equation}
\end{align}
where the state vector ${\bf{x}}_i(t)$ is a random vector in $\mathbb{R}^{n}$, which can contain the velocity, acceleration or angular velocity, etc., and it depends on the way the mobilities of nodes are modelled.
The additive uncertainties ${\bf{w}}_i(t) \in \mathbb{R}^{n}$ (assumed to be a second-order moment process) can represent the airflow for aircrafts (see~\secref{sec:Uniform Circular Motions}), velocity uncertainty for mobile phone users, etc.
The location of node $i$ in the $d$-dimensional space is denoted as ${\bf{y}}_i(t) \in \mathbb{R}^{d}$.
The constant matrices ${\bf{A}}_i \in \mathbb{R}^{n \times n}$, and ${\bf{C}}_i \in \mathbb{R}^{d \times n}$ are the model parameters determined by the node dynamics.
The initial vector for the differential equation~\eqref{eqn:state-equation} is ${\bf{x}}_i(s)$ at time $s$ and is independent from ${\bf w}_i(t),~\forall t > s$, because ${\bf{w}}_i(t)$ only affects the future behavior of ${\bf x}_i(t)$.

For the UCM example in \secref{sec:Uniform Circular Motions}, we have
\begin{align}
\renewcommand*{\arraystretch}{0.7}
\label{eqn:ucm-params}
	{\bf{A}}_i = \begin{bmatrix}
	0 & \omega_{i}\\
	-\omega_{i} & 0
	\end{bmatrix},\quad
	{\bf{C}}_i = \begin{bmatrix}
	1 & 0\\
	0 & 1
\end{bmatrix},
\end{align}
and indeed ${\bf{y}}_i(t) = {\bf{x}}_i(t)$.
This shows UCM is a second-order linear mobility model.

\begin{definition}[Homogenous Mobility]\label{def:HomogenousInterfererMobility}
If pairs $({\bf A}_i,{\bf C}_i)$ for all the nodes are equal, i.e.,
\begin{align}
	{\bf A}_i = {\bf A},\quad{\bf C}_i = {\bf C}, \quad i \in \{1,\dotsc,N\}
\end{align}
and ${\bf w}_i(t)$ for all nodes are i.i.d., then the node mobilities are homogenous.
\end{definition}
%

\begin{remark}
The random walk and discrete-time Brownian motion models given in \cite{GongZ2010,GongZ2014} are homogenous and can be regarded as a special case when~\eqref{eqn:state-equation} is discretized and ${\bf{A}} = 0$.  Thus, they are homogenous first-order linear mobility models.
The one-dimensional homogenous continuous-time mobility model with
\begin{align}\label{eqn:GMparametersinHighorderLinearModel}
\renewcommand*{\arraystretch}{0.7}
{\bf{A}} = \begin{bmatrix}
0 & 1\\
0 & \ln(1-\alpha)
\end{bmatrix},\,
{\bf{C}} = \begin{bmatrix}
1 & 0
\end{bmatrix},\,
{\bf w}(t) = \begin{bmatrix}
0 \\ w^{(2)}(t)
\end{bmatrix},
\end{align}
can be discretized to recover the Gauss-Markov mobility model given in~\cite{BaiF2004}, where $\alpha\in(0,1)$ and the mean of $w^{(2)}(t)$ is the asymptotic velocity.
\end{remark}

\begin{remark}\label{rek:The reason for regarding w_i Gaussian White Noise}
For any second-order moment process, the mean vector $\Expect[\big]{{\bf{y}}_i(t)}$ and covariance matrix $\Cov[\big]{{\bf{y}}_i(t)}$ of ${\bf y}_i(t)$ can be calculated, even though the pdf of ${\bf y}_i(t)$ has no closed-form expression in general.
For a fixed covariance matrix $\Cov[\big]{{\bf{y}}_i(t)}$, the entropy is maximized when ${\bf y}_i(t)$ is Gaussian~\cite{CoverT2006}.
It implies that for all kinds of second-order moment process ${\bf w}_i(t)$, Gaussian process contains the most uncertainties.
Practically, it is difficult to determine what kind of process ${\bf w}_i(t)$ is, and hence the best choice is to conservatively (since it contains the most uncertainties) regard ${\bf w}_i(t)$ as a Gaussian process.
Furthermore, ${\bf w}_i(t)$ is independent on time $t$ in practice, we often consider it as a white noise.
Therefore, ${\bf w}_i(t)$ can be regarded as Gaussian White Noise (GWN).
Actually, GWN is widely used in modeling uncertainties like disturbances~\cite{KalmanR1961,JettoL1999,MammarellaM2008}.
In the rest of this paper, we investigate the statistics of ${\bf{y}}_i(t)$ when ${\bf w}_i(t)$ is GWN.
It can be proved that, if ${\bf w}_i(t)$ is Gaussian in~\eqref{eqn:state-equation}, then ${\bf{y}}_i(t)$ in~\eqref{eqn:output-equation} is still Gaussian.
Thus, the pdf of ${\bf{y}}_i(t)$ can be determined by its mean and variance.
\end{remark}

\begin{lemma}[PDF of Node Distribution]\label{lem:PDFofNodeDistributions}
Assume that ${\bf w}_i(t)$ is GWN, the pdf of ${\bf{y}}_i(t)$ at time $t$ is a Gaussian distribution with parameters
\begin{align}
	\Expect[\big]{{\bf{y}}_i(t)}
		&= {\bf C}_i e^{{\bf A}_i(t-s)}{\bf{x}}_i(s),\label{eqn:MeanofNodeDistributions} \\
	\Cov[\big]{{\bf{y}}_i(t)}
		&= {\bf C}_i \Theta_{x_i}(t) {\bf C}_i^T,\label{eqn:VarianceofNodeDistributions}
\end{align}
where
\begin{align}\label{eqn:ParametersinVarianceofNodeDistributions}
\Theta_{x_i}(t) = \int_{s}^t e^{{\bf A}_i(t-\tau)} \Cov{{\bf w}_i(\tau)}\, e^{{\bf A}_i^T(t-\tau)}\,\mathrm{d}\tau.
\end{align}
\end{lemma}

\begin{IEEEproof}
See Appendix~\emph{\ref{app:Proof of PDF of Node Distributions}}.
\end{IEEEproof}

\subsection{Dynamic Reference Point}

A static reference point can be used when we investigate the interference statistics at a fixed point, like at a fixed base station.
However, if we want to analyze the interference statistics of a mobile node in a MANET, e.g., a moving robot or a UAV (see Fig.~\ref{fig:UAVsScanInterference}), the reference point should be dynamic with possible uncertainties.
%

We assume the dynamic reference point, denoted as ${\bf y}_0(t)$, satisfies
\begin{align}
	{\dot{\bf x}}_0(t)
		&= {\bf{A}}_0{\bf x}_0(t) + {\bf w}_0(t)\label{eqn:state-equation-0} \\
	{\bf{y}}_0(t)
		&= {\bf{C}}_0{\bf x}_0(t),\label{eqn:output-equation-0}
\end{align}
and the relative location of node $i$ from this reference point is given by
\begin{align}
\label{eqn:LocationfromReferencePoint}
	\overline{\bf y}_i(t) := {\bf y}_i(t) - {\bf y}_0(t).
\end{align}
Let $f_y\left({\bf y}_i(t)\right)$ denote the pdf of ${\bf y}_i(t)$.  The following Lemma derives the pdf of $\overline{\bf y}_i(t)$, denoted by $f_y\left(\overline{\bf y}_i(t)\right)$.

\begin{lemma}[Node Distribution w.r.t. a Dynamic Reference Point]\label{lem:NodeDistributionunderDynamicReferencePoint}
Assuming the mobility model of nodes and reference point are GLC with GWN, the location of the $i$th node relative to the dynamic reference point is Gaussian distributed at time $t$ with mean
\begin{equation}
	\Expect[\big]{\overline{\bf{y}}_i(t)}
		= {\bf C}_i e^{{\bf A}_i(t-s)}{\bf{x}}_i(s) - {\bf C}_0
		e^{{\bf A}_0(t-s)}{\bf{x}}_0(s)
		\label{eqn:MeanofGaussianNodeDistributionsunderReferencePoint}
\end{equation}
and variance
\begin{equation}
	\Cov[\big]{\overline{\bf{y}}_i(t)}
		= {\bf C}_i \Theta_{x_i}(t) {\bf C}_i^T + {\bf C}_0 \Theta_{x_0}(t){\bf C}_0^T.
		\label{eqn:VarianceofGaussianNodeDistributionsunderReferencePoint}
\end{equation}
\end{lemma}

\begin{IEEEproof}
\lemref{lem:NodeDistributionunderDynamicReferencePoint} follows from \lemref{lem:PDFofNodeDistributions} and \eqref{eqn:LocationfromReferencePoint}.
\end{IEEEproof}

\section{Interference Prediction}\label{sec:InterferencePredictions}

The main problem of study in this work is to characterize the interference received at a reference point at a future time instant given the interferers' mobility and location information at the current time instant, i.e., interference prediction from the current time into the future.
We use the GLC mobility model defined in the previous section to describe the mobility of the interferers and the reference node.
Specifically, the quantity ${\bf y}_i(t)$ $i\in\{0,\ldots,N\}$ in~\eqref{eqn:output-equation} or~\eqref{eqn:output-equation-0} is the random variables that describe the locations of nodes (interferers or reference node) at time $t$ with a known initial condition at time $s$, where $t$ can be viewed as a future time instant and $s$ can be viewed as the current time instant.
To make the time-dependency more explicit, we rewrite it as ${\bf y}_i(t|s)$ in the remainder of the paper.

\subsection{Problem Description}\label{sec:ProblemDescriptions}

Suppose reference node $0$ is receiving information from its transmitter.
Assume that there are $N$ mobile interferers in this network whose mobilities are modeled by~\eqref{eqn:state-equation} and~\eqref{eqn:output-equation}, and their interference lasts for time duration $[t_0, t_f]$.
The aggregate interference on the reference node at time $t\in[t_0,t_f]$, conditioned on knowing the interferers' node dynamics and locations at time $s$ with $s\leq t$, can be defined as
\begin{align}\label{eqn:TotalInterference}
     I(t|s) = \sum_{i=1}^{N} h_i \, g\big(\|\overline{\bf y}_i(t|s)\|\big),
\end{align}
where $h_i$ is the multipath fading gain with $\Expect{h_i} = 1$, which is independent of $\overline{\bf y}_i(t|s)$ and time $t$.
$g(\cdot)$ denotes the path loss function, and $\|\cdot\|$ is the Euclidean norm.
For the general case of $s < t$, the quantity $I(t|s)$ represents the interference prediction at a future time instant $t$ based on the information available at the current time instant $s$, and hence, it is a random variable due to the uncertainty in the mobility over the time duration from $s$ to $t$.
For the very special case of $s=t$, the quantity $I(t|t)$ represents the actual interference received at time $t$ which is a constant instead of a random variable.
In fact, $I(t|t)$ can be viewed as a realization of the random variable $I(t|s)$ for $s < t$.
In this paper, we simply refer to $I(t|s)$ as interference prediction.

We consider the problem of predicting the statistics of interference received by reference node $0$ at time $t$ with the available information from current time $s\leq t$.
Denote
\begin{align}\label{eqn:Its}
	\Stat[\big]{I(t|s)}
		= \Stat[\Big]{\sum_{i=1}^{N} h_i g\big(\|\overline{\bf y}_i(t|s)\|\big)},
\end{align}
where $\Stat{\cdot}$ can be any statistics of the interference prediction, such as the mean value, and $\overline{\bf y}_i(t|s)$ is a conditioned random vector which represents the relative location of interferer $i$ from node $0$.
In most cases, evaluating $\Stat[\big]{I(t|s)}$ requires the pdf of $\overline{\bf y}_i(t|s)$, which can be determined by \lemref{lem:NodeDistributionunderDynamicReferencePoint}.

In the remainder of this section, the mean value and MGF of the interference prediction will be considered in \secref{sec:PredictionforMeanInterference} and \secref{sec:PredictionforInterferenceMomentGeneratingFunction}, respectively.
In \secref{sec:GaussianPointProcessFunctionalandItsSeriesForm}, we will propose a compound Gaussian point process functional as a general framework to study these statistics.
In \secref{sec:DecayofPredictedInformation:HomogenousMobilitiesandGaussianBPPs}, the information decay in interference predictions will be discussed, and the prediction has close ties to BPP modelling when time goes to infinity.

\subsection{Interference Prediction Mean}\label{sec:PredictionforMeanInterference}

The mean of the interference prediction is given by the following theorem.

\begin{theorem}[The Mean of the Interference Prediction]
\label{thm:InterferenceMeanPrediction}
The mean of the interference prediction $\Expect{I(t|s)}$ is
\begin{align}
\label{eqn:EIts}
	\Expect[\big]{I(t|s)} \!=\! \sum_{i=1}^{N}\int_{\mathbb{R}^d} \!\!g\big(\|\overline{\bf y}_i(t|s)\|\big)
	 f_y\left(\overline{\bf y}_i(t|s)\right) \mathrm{d}(\overline{\bf y}_i(t|s)),
\end{align}
where $f_y\big(\overline{\bf y}_i(t|s)\big)$ is the pdf of $\overline{\bf y}_i(t|s)$, and can be obtained from \lemref{lem:NodeDistributionunderDynamicReferencePoint}.
\end{theorem}

\begin{IEEEproof}
The mean value of interference prediction is
\begin{align}\label{eqn:EItsProof1}
	\Expect[\big]{I(t|s)} = \sum_{i=1}^{N}\Expect[\big]{I_i(t|s)}.
\end{align}
According to~\eqref{eqn:TotalInterference}, $\mathbb{E}[h_i] = 1$ and the independence of $h_i$ and $\overline{\bf y}_i(t|s)$, we can derive
\begin{align}\label{eqn:EItsProof2}
	\Expect[\big]{I_i(t|s)} = \int_{\mathbb{R}^d} g(\|\overline{\bf y}_i(t|s)\|)\, f_y\left(\overline{\bf y}_i(t|s)\right) \mathrm{d}\left(\overline{\bf y}_i(t|s)\right).
\end{align}
Thus,~\eqref{eqn:EIts} can be obtained.
\end{IEEEproof}


The calculation for the mean of the interference prediction will be discussed in \secref{sec:GaussianPointProcessFunctionalandItsSeriesForm}.

\subsection{Interference Prediction MGF}\label{sec:PredictionforInterferenceMomentGeneratingFunction}

Similar to the mean interference, the MGF of the interference prediction can be derived in the following theorem.

\begin{theorem}[The MGF of the Interference Prediction]\label{thm:InterferenceMomentGeneratingFunctionPrediction}
The MGF of the interference prediction is
\begin{equation}
\label{eqn:MFGIts}
	\Expect[\big]{e^{\beta I(t|s)}} \!= \!\prod_{i=1}^N \!\int_{\mathbb{R}^d} \!\!\!\!\mathbb{E}_h\big[e^{\beta h_i g(\|\overline{\bf y}_i(t|s)\|)}\big] f_y\big(\overline{\bf y}_i(t|s)\big) \mathrm{d}\big(\overline{\bf y}_i(t|s)\big),
\end{equation}
where $f_y\big(\overline{\bf y}_i(t|s)\big)$ is the pdf of $\overline{\bf y}_i(t|s)$, and can be obtained from \lemref{lem:NodeDistributionunderDynamicReferencePoint}.
\end{theorem}

\begin{IEEEproof}
With~\eqref{eqn:TotalInterference} and independence of $h_i$ and $\overline{\bf y}_i(t)$, \thmref{thm:InterferenceMomentGeneratingFunctionPrediction} can be proved.
\end{IEEEproof}

If the power fading is Nakagami-$m$ ($m = 1$ gives the Rayleigh fading), i.e.,
\begin{align}\label{eqn:Nakagami-mFading}
f_h(x) = \frac {m^m x^{(m-1)} e^{-mx}} {\Gamma(m)},
\end{align}
then the MGF of the interference prediction can take a more specific form stated as follows.

\begin{corollary}[The MGF of the interference prediction w.r.t. Nakagami-m Fading]\label{cor:InterferenceMomentGeneratingFunctionPredictionunderNakagami-mFading}
With Nakagami-$m$ Fading~\eqref{eqn:Nakagami-mFading}, $\Expect{e^{\beta I(t|s)}}$ in~\eqref{eqn:MFGIts} becomes
\begin{align}
\label{eqn:MFGItsNakagami-mFading}
	\prod_{i=1}^N \int_{\mathbb{R}^d} \Big[\frac {m} {m - \beta g\big(\|\overline{\bf y}_i(t|s)\|\big)}\Big]^{m} f_y\left(\overline{\bf y}_i(t|s)\right) \mathrm{d}\left(\overline{\bf y}_i(t|s)\right),
\end{align}
where $\beta\, g(\|\overline{\bf y}_i(t)\|) < 1$.
\end{corollary}

\begin{remark}
As already discussed, $I(t|s)$ is a random variable that represents the interference prediction at a future time instant $t$ based on the information available at the current time instant $s$.
The uncertainty of the random variable can be computed using its MGF.
For instance, one can compute how much the realizations of $I(t|s)$ deviates from its mean value using the variance
\begin{align}
\mathrm{Var}[I(t|s)] = \Expect{I^2(t|s)} - \left(\Expect{I(t|s)}\right)^2,
\end{align}
where the first and second moments of $I(t|s)$ are used.
Since the actual interference received at time $t$, i.e., $I(t|t)$, is a realization of the interference prediction $I(t|s)$, the variance computed above tells on average how much the actual interference received at time $t$ deviates from the predicted value at time $s$ using the mean prediction.
\end{remark}

The calculation for the MGF of the interference prediction is discussed in \secref{sec:GaussianPointProcessFunctionalandItsSeriesForm}.

Although the expressions of either the mean or MGF of the interference prediction do not usually admit any closed form due to the generality of the GLC mobility model, exceptions are found in some special cases where closed-form expressions are obtained.
These results are discussed in Remark~\ref{rek:Condition for Some Closed-forms} in the next section.

\subsection{Compound Gaussian Point Process Functional and its Series Form} \label{sec:GaussianPointProcessFunctionalandItsSeriesForm}

In both \secref{sec:PredictionforMeanInterference} and \secref{sec:PredictionforInterferenceMomentGeneratingFunction}, the mean and MGF have similar integrals to evaluate.
This suggests there may be some general methods to compute these quantities.
Here, we define the Compound Gaussian Point Process Functional (CGPPF) as a general framework for computing these statistics.
Its series form is also given, which will also be useful in analyzing a limiting property of interference prediction in \secref{sec:DecayofPredictedInformation:HomogenousMobilitiesandGaussianBPPs}.

\begin{definition}[Compound Gaussian Point Process Functional]\label{def:CompoundGaussianPointProcessFunctional}
The CGPPF is a functional $G\colon \mathcal{V}\rightarrow\mathbb{R}$ of the form
\begin{align}\label{eqn:GaussianPointProcessFunctional}
G[\nu] = \mathbb{E}_y \big[\nu\big(\|{\bf y}\|\big)\big] = \int_{\mathbb{R}^d} \nu\big(\|{\bf y}\|\big)\,f_y\big({\bf y}\big)\,\mathrm{d}{\bf y},
\end{align}
where $\nu\in\mathcal{V}$ is a Lebesgue Integrable function, and $f_y\big({\bf y}\big)$ is a pdf of $d$-dimensional Gaussian distribution given by
\begin{align}
	f_y\big({\bf y}\big) = \frac {1} {(2\pi)^{\frac {d} {2}}|{\bf \Sigma}|^{\frac {1} {2}}} e^{-\frac {1} {2} ({\bf y} - {\boldsymbol \mu})^T {\bf \Sigma}^{-1} ({\bf y} - {\boldsymbol \mu})},
\end{align}
where ${\boldsymbol \mu} \in \mathbb{R}^d$ and ${\bf \Sigma} \in \mathbb{R}^{d\times d}$ are mean vector and covariance matrix of location vector ${\bf y}$.
For example, ${\boldsymbol \mu} = \Expect[\big]{\overline{\bf{y}}_i(t|s)}$ and ${\bf \Sigma} = \Cov[\big]{\overline{\bf{y}}_i(t|s)}$ are mean and covariance of $\overline{\bf{y}}_i(t|s)$.
\end{definition}

\begin{remark}
For interferer $i$ and dynamic reference point $0$, the following are derived:
If $\nu(\|\cdot\|) = g(\|\overline{\bf y}_i(t|s)\|)$, then~\eqref{eqn:GaussianPointProcessFunctional} returns the mean of the interference prediction from interferer $i$.
If $\nu(\|\cdot\|) = \mathbb{E}_h\left[e^{\beta h_i(t) g(\|\overline{\bf y}_i(t|s)\|)}\right]$, then~\eqref{eqn:GaussianPointProcessFunctional} gives the MGF of the interference prediction from interferer $i$.
Note that the path loss function, $g(\cdot)$, can be arbitrary.
\end{remark}

\begin{remark}\label{rek:Condition for Some Closed-forms}
In most cases, this functional cannot be simplified to a closed-form expression, and numerical integration is needed.
Nevertheless, if
\begin{align}\label{eqn:Condition for Some Closed-forms}
{\boldsymbol \mu} = {\bf 0},~{\bf \Sigma} = \operatorname{diag}\{\sigma,\dotsc,\sigma\}
\end{align}
is satisfied, we can get closed-form expressions for the first and second moments of the interference prediction, where $\sigma > 0$ is the std (standard deviation) of all components in location vector ${\bf y}$.
These expressions are given in Appendix~\ref{app:Some Closed Forms for CGPPF}.
The usefulness of condition~\eqref{eqn:Condition for Some Closed-forms} will be further discussed in \secref{sec:DecayofPredictedInformation:HomogenousMobilitiesandGaussianBPPs}.
\end{remark}

If the integral in~\eqref{eqn:GaussianPointProcessFunctional} exists, the CGPPF can be expanded into a series form, which is the cornerstone for analyzing the limit properties of interference predictions in \secref{sec:DecayofPredictedInformation:HomogenousMobilitiesandGaussianBPPs}.

\begin{theorem}[Series Form for Compound Gaussian Point Process Functional]\label{thm:SeriesFormforGaussianPointProcessFunctional}
Let ${\bf P}$ be an orthogonal matrix such that ${\bf P}^{T}$\!{\boldmath${\mu}$} $ = $ {\boldmath ${\bf \eta}$} $=[\eta_i]_{d\times1}$ and ${\bf P}^{T}{\bf\Sigma}^{-1}{\bf P}=\operatorname{diag}\{1/\sigma_1^2,\dotsc,1/\sigma_d^2\}$, then $G[\nu]$ can be written as
\begin{align}\label{eqn:Gnu}
	\frac {1} {(2\pi)^{\frac {d} {2}}|{\bf \Sigma}|^{\frac {1} {2}}} \lim_{R\rightarrow\infty}\sum_{n=0}^{\infty} \frac {(-1)^n} {2^nn!} \sum_{k_1+k_2+k_3=n}
	\binom{n}{k_1,k_2,k_3}
	\Omega\,\Psi[\nu].
\end{align}
The parameters of~\eqref{eqn:Gnu} are listed as follows:
\begin{align}
\Psi[\nu] &= \int_0^{R} \nu\left(r\right)r^{2k_1+k_2+d-1} \mathrm{d}r,\label{eqn:Psi}\\
\begin{split}
\Omega &= \Big[\sum_{a=1}^{d} \Big(\frac {\eta_{a}} {\sigma_a}\Big)^{2} \Big]^{k_3}\cdot\\&\!\!\!\!\!\!\sum_{k_{1}^{(1)}+\dotsb+k_{1}^{(d)}=k_1 \atop k_{2}^{(1)}+\dotsb+k_{2}^{(d)}=k_2}
\binom{k_1}{k_{1}^{(1)},\dotsc,k_{1}^{(d)}}\binom{k_2}{k_{2}^{(1)},\dotsc,k_{2}^{(d)}}\,\Xi,\label{eqn:Omega}
\end{split}\\
\begin{split}
\Xi &= \prod_{1\leq a\leq d \atop 1\leq b\leq d} \left\{\Big(\frac {1} {\sigma_a}\Big)^{2k_{1}^{(a)}} \Big(-2\frac{\eta_{b}}{\sigma_b^2}\Big)^{k_{2}^{(b)}}\cdot\right.\\ &~~~~~~~~~~~~~~~~~~~\left.\int_{\bf \Theta} (\Phi_{a})^{2k_{1}^{(a)}} (\Phi_{b})^{k_{2}^{(b)}} V({\boldsymbol \phi})\, \mathrm{d}{\boldsymbol \phi}\right\},\label{eqn:Xi}
\end{split}
\end{align}
where the integration is with respect to the $d-1$ dimensional vector ${\boldsymbol \phi} = \Big[\phi_1, \phi_2, \dotsc, \phi_{d-1}\Big]^T$ over the domain ${\bf \Theta} = \big\{{\boldsymbol \phi}:{\boldsymbol \phi} \in \underbrace{[0,~\pi]\times\cdots\times[0,~\pi]}_{d-2}\times[0,~2\pi]\big\}$, and
\begin{align}
\begin{split}
[\Phi_{1},\ldots,\Phi_{d}]^T &= \Big[\cos\phi_1, \sin\phi_1\cos\phi_2, \dotsc,\\ &~~~~~~~~\prod_{p=1}^{d-2}\sin\phi_{p}\cos\phi_{d-1}, \prod_{p=1}^{d-1}\sin\phi_{p}\Big]^T,\label{eqn:Phi}
\end{split}\\
V({\boldsymbol \phi})\,\mathrm{d}{\boldsymbol \phi} &= \prod_{q=1}^{d-1}(\sin\phi_l)^{d-q-1}\mathrm{d}\phi_q.\label{eqn:ThetaVolumeElement}
\end{align}
\end{theorem}

\begin{IEEEproof}
See Appendix~\ref{app:ProofofGaussianPointProcessFunctionalSeriesForm}.
\end{IEEEproof}

\begin{remark}
It should be noted that there are two required integrations in Theorem~\ref{thm:SeriesFormforGaussianPointProcessFunctional}.
For~\eqref{eqn:Psi}, it has a closed-form expression when calculating the mean and MGF of the interference predictions with the widely-used path-loss function of the form
\begin{align}\label{eqn:PathLossFunction}
	g\big(\|{\bf y}\|\big) =  \frac {1} {\epsilon + \|{\bf y}\|^{\alpha}}
\end{align}
where $\epsilon \geq 0$ (here $\epsilon = 0$ refers to a singular path loss), and $\alpha$ is the path-loss exponent.
The expression of~\eqref{eqn:Psi} is given in Appendix~\ref{app:IntegrationsinGaussianPointProcessFunctionalSeriesForm}.

To evaluate $\Omega$ in~\eqref{eqn:Omega}, we need to compute the integral in~\eqref{eqn:Xi}, i.e.,
\begin{align}\label{eqn:Integral of Theta}
	\int_{\bf \Theta} (\Phi_{a})^{2k_{1}^{(a)}} (\Phi_{b})^{k_{2}^{(b)}} V({\boldsymbol \phi})\,\mathrm{d}{\boldsymbol \phi},
\end{align}
the calculation of which is rather complex.
However, in the real world, $1\leq d\leq 3$, and it is relatively easy to deal with.
In Appendix~\ref{app:IntegrationsinGaussianPointProcessFunctionalSeriesForm}, we give the closed-form expression of~\eqref{eqn:Omega} for the cases of $d = 1$ (e.g., a vehicular network on a highway) and $d = 2$ (which is the most common scenario of interest).
\end{remark}

\subsection{Gaussian BPP Approximations for Interference Prediction}\label{sec:DecayofPredictedInformation:HomogenousMobilitiesandGaussianBPPs}

The interference prediction method discussed in previous subsections is naturally based on the mobility model of individual interferers.
Although prior work on interference analysis for MANETs did not explicitly study the interference prediction problem, one useful idea from them is to use a certain point process to approximate the node distribution in the network from which the time-invariant interference statistics can be derived.
For networks where its node distribution (in distant future) can be well approximated using a point process, the information on the initial positions of nodes and node mobilities becomes unnecessary in determining the interference statistics.
Clearly, not all mobile networks can have such a nice property.
In this subsection, we study the condition under which the interference prediction at a time of far future can be well approximated based on a simple point process.
We will see that the series form of the CGPPF will be useful in determining such a condition.

Firstly we give the definition of Gaussian Binomial Point Process (Gaussian BPP).

\begin{definition}[Gaussian BPP]\label{def:GaussianBPP}
Let $f_{\bf y}$ be a pdf of Gaussian distribution with support $\mathbb{R}^d$. A Gaussian BPP with $N$ points on $\mathbb{R}^d$ is a set of i.i.d. random vectors $\{{\bf y}_1,\dotsc,{\bf y}_N\}$, each with pdf $f_{\bf y}$.
\end{definition}

Due to the effect of Gaussian white noise, i.e., ${\bf w}_i(t)$ in~\eqref{eqn:state-equation}, the uncertainty of our prediction will increase with time, which implies the information available for prediction will decay.
In this section, we will focus on the prediction into the far future with $N$ interferers governed by homogenous mobilities.
The following theorem gives a necessary and sufficient condition that the statistics of interference predictions can be approximated as those generated by interferers whose locations follow a Gaussian BPP when $t$ becomes large enough.

\begin{theorem}[Necessary and Sufficient Condition for Gaussian BPP Approximation]\label{thm:NecessaryandSufficientConditionforGaussianBPPs}
Assume that all interferers' mobilities are homogenous, as defined in Definition~\ref{def:HomogenousInterfererMobility}, and the integral in~\eqref{eqn:GaussianPointProcessFunctional} exists, $\forall \nu \in \mathcal{V}$, the predictions satisfy
\begin{align}\label{eqn:AsymptoticallyConvergetoaBPP}
\lim_{t\rightarrow\infty} (G_i[\nu] - G_j[\nu]) = 0,~\forall i,j \in \left\{1,\ldots,N\right\}
\end{align}
if and only if
\begin{multline}\label{eqn:NecessaryandSufficientCondtionforGaussianBPPs}
\left[\lim_{t\rightarrow\infty} \frac {\eta_a^{(i)}} {\sigma_a^{(i)}} - \lim_{t\rightarrow\infty} \frac {\eta_a^{(j)}} {\sigma_a^{(j)}}\right] = 0,~\\
\forall a = \{1,\dotsc,d\},~\forall i,j \in \left\{1,\ldots,N\right\}
\end{multline}
holds, where $\eta_a^{(i)}$ and $\sigma_a^{(i)}$ are defined in \thmref{thm:SeriesFormforGaussianPointProcessFunctional} with superscripts $(i)$ for $i$th interferer and
\begin{align}\label{eqn:Gnui}
\begin{split}
G_i[\nu] &= \int_{\mathbb{R}^d} \nu(\|\overline{\bf y}_i\|) \frac {1} {(2\pi)^{\frac {d} {2}}|{\bf \Sigma}_i|^{\frac {1} {2}}} e^{-\frac {1} {2} \left(\overline{\bf y}_i - {\boldsymbol \mu}_i\right)^T {\bf \Sigma}_i^{-1} \left(\overline{\bf y}_i - {\boldsymbol \mu}_i\right)} \mathrm{d}\overline{\bf y}_i.
\end{split}
\end{align}
In~\eqref{eqn:Gnui}, ${\boldsymbol \mu}_i$, ${\bf \Sigma}_i$ and $\overline{\bf y}_i$ stand for $\Expect[\big]{\overline{\bf{y}}_i(t|s)}$, $\Cov[\big]{\overline{\bf{y}}_i(t|s)}$ and $\overline{\bf y}_i(t|s)$, respectively.
\end{theorem}

\begin{IEEEproof}
If the mobilities of all interferers are homogenous, from \lemref{lem:NodeDistributionunderDynamicReferencePoint}, the covariance matrix ${\boldsymbol \Sigma}_i$ in~\eqref{eqn:Gnui} for all the interferers are the same.
Thus, if the integral in~\eqref{eqn:GaussianPointProcessFunctional} exists, $\lim\limits_{t\rightarrow\infty} \big\{G_i[\nu] - G_j[\nu]\big\}$ can be written as
\begin{align}\label{eqn:Subtraction}
\begin{split}
&\frac {1} {(2\pi)^{\frac {d} {2}}|{\bf \Sigma}_1|^{\frac {1} {2}}} \lim_{R\rightarrow\infty}\sum_{n=1}^{\infty} \Bigg\{\frac {(-1)^n} {2^nn!}\cdot \\ &~~~~~~~\sum_{k_1+k_2+k_3=n\atop k_3\neq0}
\binom{n}{k_1,k_2,k_3}
\lim_{t\rightarrow\infty}\left(\Omega_i - \Omega_j\right) \Psi[\nu]\Bigg\}.
\end{split}
\end{align}
Obviously, formula~\eqref{eqn:Subtraction} equals $0$ if and only if~\eqref{eqn:AsymptoticallyConvergetoaBPP} holds.

\emph{Necessity}. By the contrapositive, if the~\eqref{eqn:NecessaryandSufficientCondtionforGaussianBPPs} does not hold, $\lim\limits_{t\rightarrow\infty}\left(\Omega_i - \Omega_j\right) \neq 0$ (since arbitrary $\nu$ determines arbitrary~\eqref{eqn:Integral of Theta} in~\eqref{eqn:Xi}, if $\lim\limits_{t\rightarrow\infty}\left(\Omega_i - \Omega_j\right) = 0$,~\eqref{eqn:NecessaryandSufficientCondtionforGaussianBPPs} must be satisfied).
As a result, the~\eqref{eqn:Subtraction} does not equal $0$.
Therefore, the contrapositive is established, and the necessity of~\eqref{eqn:NecessaryandSufficientCondtionforGaussianBPPs} is proved.

\emph{Sufficiency}. If~\eqref{eqn:NecessaryandSufficientCondtionforGaussianBPPs} is satisfied, then the~\eqref{eqn:Subtraction} equals $0$, so as for~\eqref{eqn:AsymptoticallyConvergetoaBPP}.
Thus the sufficiency of~\eqref{eqn:NecessaryandSufficientCondtionforGaussianBPPs} is established.
\end{IEEEproof}

\begin{remark}
\thmref{thm:NecessaryandSufficientConditionforGaussianBPPs} tells us the interference predictions, from interferers with homogenous mobility asymptotically converge to the predictions from a Gaussian BPP.
Nevertheless, if~\eqref{eqn:NecessaryandSufficientCondtionforGaussianBPPs} does not hold, we cannot use the BPP approximation.
In \secref{sec:SimulationExamples}, we will see examples in both scenarios.
It should be noted that $\eta_a^{(i)}$ and $\sigma_a^{(i)}$ in~\eqref{eqn:NecessaryandSufficientCondtionforGaussianBPPs} can be easily calculated for a given GLC mobility model (see \secref{sec:SimulationExamples} for examples).
\end{remark}

\begin{remark}\label{rek:NecessaryandSufficientConditionforGaussianBPPs}
Gaussian BPP approximation implies that the mean and MGF of the interference prediction become
\begin{align}
	\Expect[\big]{I(t|s)} &\approx N\, G_i\big[g\big(\|\overline{\bf y}_i(t|s)\|\big)\big],\quad\forall i\label{eqn:Interference Mean Approximation} \\
	\Expect[\big]{e^{\beta I(t|s)}} &\approx G_i\Big[\Big(\frac {m} {m - \beta g(\|\overline{\bf y}_i(t|s)\|)}\Big)^{m}\Big]^N\!\!\!,\quad\forall i,\label{eqn:Interference MGF Approximation}
\end{align}
when $t\gg s$.
%
%
Therefore, we can calculate the mean or MGF of the interference prediction using just one (arbitrary) interferer's information instead of all interferers, and the computation is significantly simplified.
Note that~\eqref{eqn:Interference Mean Approximation} and~\eqref{eqn:Interference MGF Approximation} change with time $t$, because the distribution of the predicted location $\overline{\bf y}(t|s)$ varies with $t$.
\end{remark}

\begin{corollary}[Necessary and Sufficient Condition for Gaussian BPP Approximation with ${\boldsymbol \mu} = {\bf 0}$]\label{cor:NecessaryandSufficientConditionforGaussianBPPs0}
Assume that all interferers' mobilities are homogenous and the integral in~\eqref{eqn:GaussianPointProcessFunctional} exists, $\forall \nu \in \mathcal{V}$, the predictions satisfy
\begin{align}\label{eqn:AsymptoticallyConvergetoaBPPo}
\lim_{t\rightarrow\infty} (G_i[\nu] - G_o[\nu]) =  0,~\forall i \in \left\{1,\ldots,N\right\}
\end{align}
if and only if
\begin{align}\label{eqn:NecessaryandSufficientCondtionforGaussianBPPso}
\lim_{t\rightarrow\infty} \frac {\eta_a^{(i)}} {\sigma_a^{(i)}} = 0,~\forall a \in \{1,\dotsc,d\},~\forall i \in \left\{1,\ldots,N\right\}
\end{align}
holds, and
\begin{align}\label{eqn:Gnuo}
G_o[\nu] = \int_{\mathbb{R}^d} \nu(\|{\bf y}\|) \frac {1} {(2\pi)^{\frac {d} {2}}|{\bf \Sigma}|^{\frac {1} {2}}} e^{-\frac {1} {2} {\bf y}^T {\bf \Sigma}^{-1} {\bf y}} \mathrm{d}{\bf y},
\end{align}
where ${\boldsymbol \mu} = {\boldsymbol \mu}_1 = \ldots = {\boldsymbol \mu}_N$ (${\boldsymbol \mu}_i = \Expect[\big]{\overline{\bf{y}}_i(t|s)}$) and ${\bf \Sigma} = {\bf \Sigma}_1 = \ldots = {\bf \Sigma}_N$ (${\bf \Sigma}_i = \Cov{\overline{\bf y}_i (t|s)}$).
\end{corollary}

\begin{remark}
Corollary~\ref{cor:NecessaryandSufficientConditionforGaussianBPPs0} implies that we can use the Gaussian BPP with ${\boldsymbol \mu} = {\bf 0}$ to approximate the mean or MGF of the interference prediction when condition~\eqref{eqn:NecessaryandSufficientCondtionforGaussianBPPso} holds.
It should be noted that it is easy to test condition~\eqref{eqn:NecessaryandSufficientCondtionforGaussianBPPso} by calculating $\eta_a^{(i)}$ and $\sigma_a^{(i)}$ for a given GLC mobility model.
Furthermore, if ${\bf \Sigma} = \operatorname{diag}\{\sigma,\dotsc,\sigma\}$ also holds, we can use the result stated in Remark~\ref{rek:Condition for Some Closed-forms} to obtain closed-form expressions for first and second moments of interference prediction given in Appendix~\ref{app:Some Closed Forms for CGPPF}.
An example of mobility model that has such nice properties is Brownian motion.
\end{remark}

\section{Simulation Examples}\label{sec:SimulationExamples}

In order to corroborate our theoretical results, simulations are presented to illustrate the effectiveness of interference prediction.
We consider three different mobile networks with the mobility model given by: (1) the basic $2$-dimensional Brownian motion, (2) $2$-dimensional Brownian motion with inertia in the velocity and (3) UCM in $3$-dimensional space.
The basic Brownian motion is a simple and commonly-used mobility model while Brownian motion with inertia is a new mobility model that has not been considered in the literature.
We will study and compare the interference prediction results for these two examples.
Specifically, we will see that the existence of inertia completely changes the limiting behavior of the interference prediction.
After that, we move from 2-dimensional examples to a 3-dimensional example of UCM which can represent the scenario of UAV target scanning similar to that in Section II.
Note that all three mobility models are special cases of the GLC mobility model developed in this work.

To make predictions, the parameters for path loss function (e.g., the path-loss exponent $\alpha$) and multipath fading (e.g., the value of $m$ for the Nakagami-m fading channel) are necessary.
In addition the following parameters are assumed to be known at time $s$:
\begin{itemize}
\item For 2D Brownian motion (\secref{sec:BrownianMotion2D}), the location for each node.
\item For 2D Brownian motion with inertia (\secref{sec:BrownianMotion2DwithInertia}), the location and velocity for each node.
\item For UCM (\secref{sec:UAVs Target Scanning 3D}), the location and angular velocity for each node.
\end{itemize}

Note that the node location and velocity at only one time instant (not frequently updated) can be obtained in many practical scenarios~\cite{AshbrookD2003,BrochJ1988,TanH2006}.
For mobile users (\secref{sec:BrownianMotion2D}), their locations can be updated by Global Positioning System (GPS)~\cite{AshbrookD2003}.
For vehicles (\secref{sec:BrownianMotion2DwithInertia}) or UAVs (\secref{sec:UAVs Target Scanning 3D}), their locations or velocities can be obtained by Differential GPS (DGPS), e.g.,~\cite{BrochJ1988,TanH2006}, and the angular velocity can be calculated by the node location, node velocity and the center location of UCM.

\subsection{2D Brownian Motion}\label{sec:BrownianMotion2D}

We consider a network having $7$ nodes with mobility governed by Brownian motion, which can be used to describe human mobility~\cite{RheeI2011}.
As a special case of our GLC mobility model, 2D Brownian motion has the parameters
\begin{align}\label{eqn:BrownianMotionParameters}
\renewcommand*{\arraystretch}{0.7}
{\bf A}_i = \begin{bmatrix}
0 & 0 \\
0 & 0
\end{bmatrix},\quad
{\bf C}_i = \begin{bmatrix}
1 & 0 \\
0 & 1
\end{bmatrix},
\end{align}
where $i = 0,1,\dotsc,6$ (that is, $6$ interferers and $1$ reference point).
Note that the reference node also does Brownian motion.
Furthermore, all the nodes (including the reference node) start moving from the origin at time $t_0 = 0$, i.e., ${\bf y}_i(0) = {\bf x}_i(0) = {\bf 0}$.

\begin{figure}
\centering
\subfigure[]{\includegraphics [width=0.72\columnwidth]{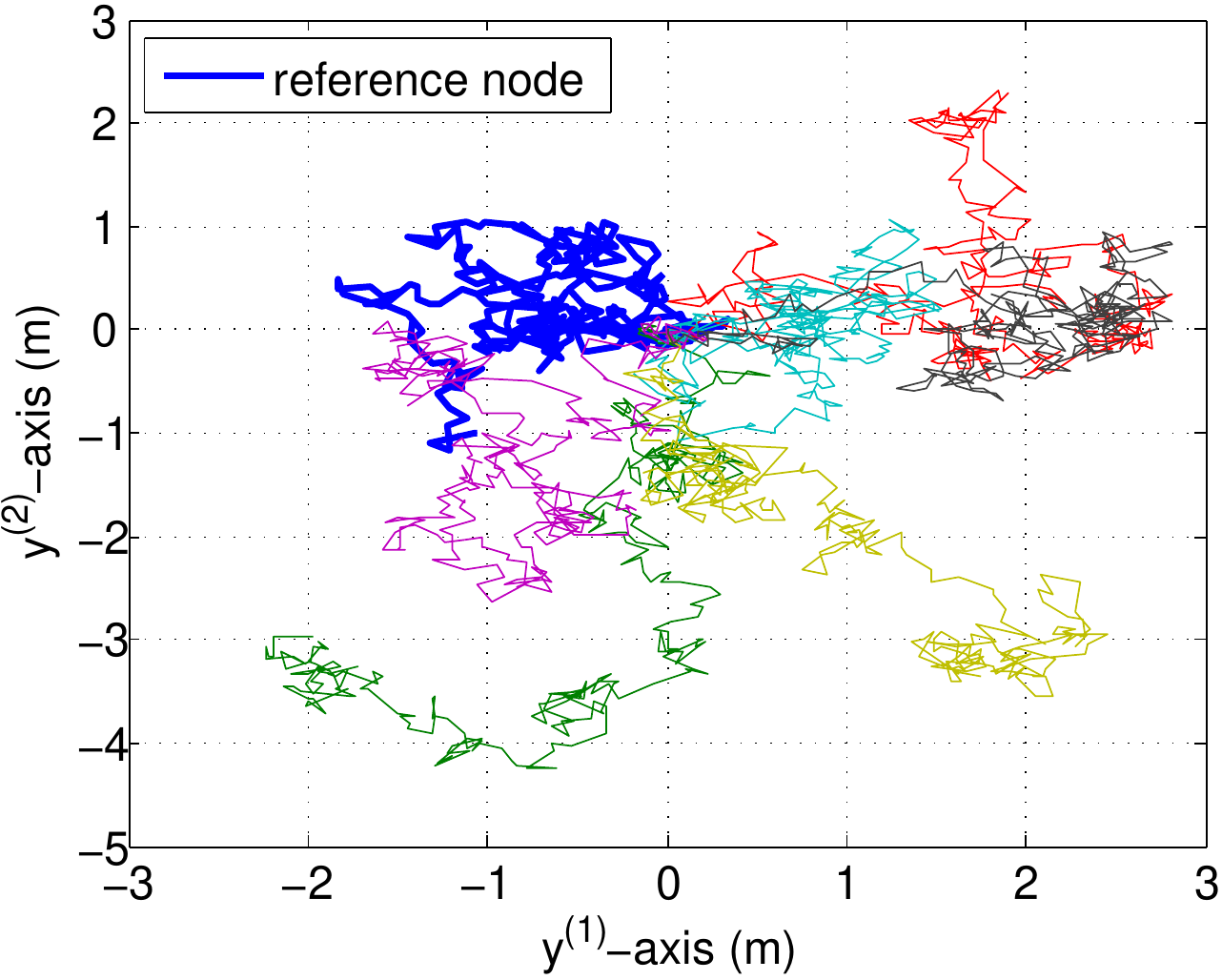}\label{fig:BrownianMotion2DLocations}}\hfill
\subfigure[]{\includegraphics [width=0.72\columnwidth]{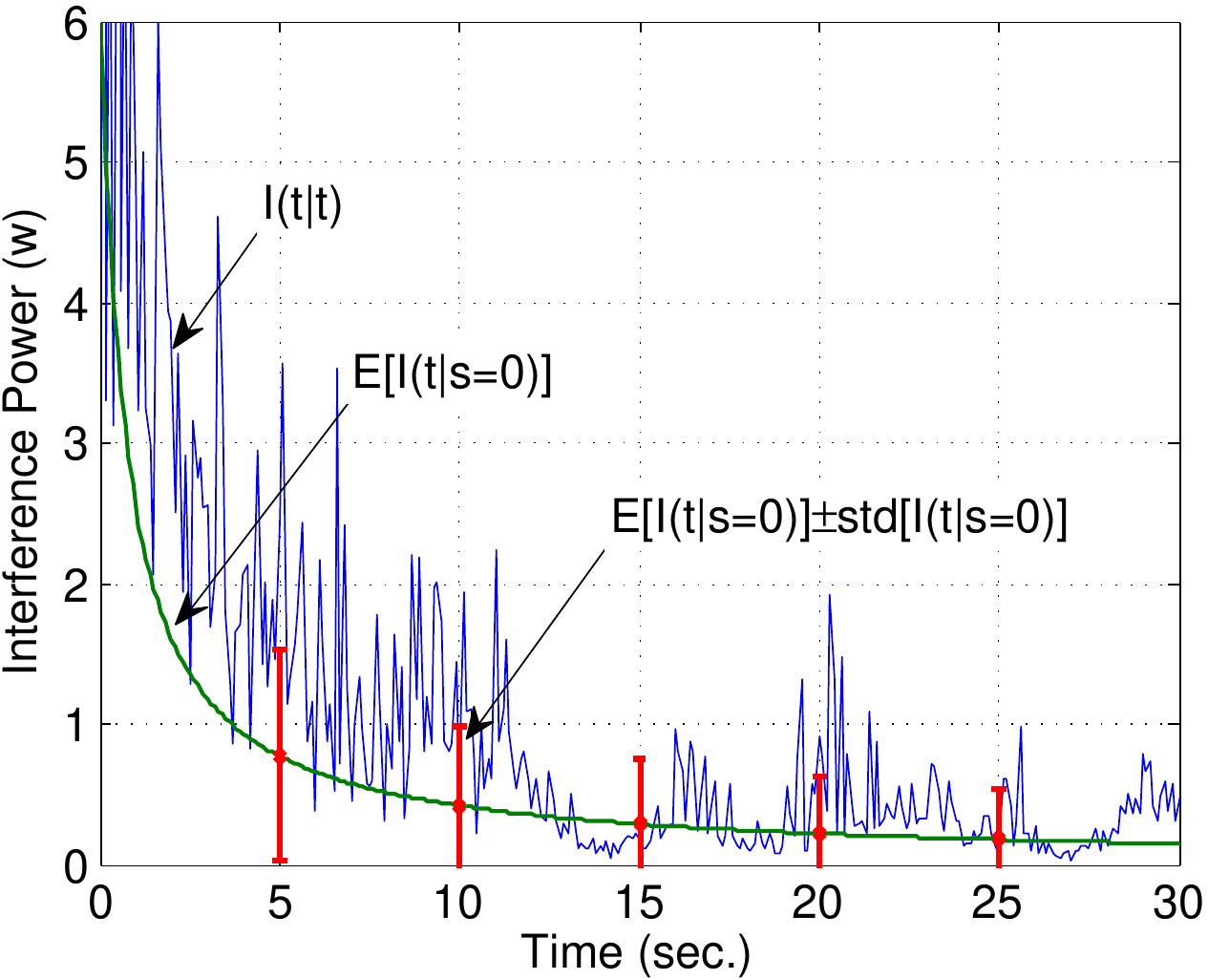}\label{fig:BrownianMotion2D_alpha4PIwithFading}}
\caption{(a) Node motions. (b) Aggregate interference at reference node and the corresponding predictions. $\Expect{I(t|s=0)}$ stands for the mean value of interference prediction based on information available at $s=0$, and $\operatorname{std}\left[I(t|s=0)\right]$ is the standard deviation of prediction. They are both dependent on time $t$ and $s$. Each error bar $\Expect{I(t|s=0)} \pm \operatorname{std}\left[I(t|s=0)\right]$ is symmetric about $\Expect{I(t|s=0)}$, and the negative part of $\Expect{I(t|s=0)} - \operatorname{std}\left[I(t|s=0)\right]$ is omitted.}
\end{figure}

The uncertainty ${\bf w}_i$ follows
\begin{align}
	{\bf w}_i(t) = \big[w_i^{(1)}(t), w_i^{(2)}(t)\big]^T,
\end{align}
where all $w_i(t)$ for all $i$ are identical GWNs with unit power.
A realization of Brownian motion is shown in Fig.~\ref{fig:BrownianMotion2DLocations}.

Considering $\epsilon = 1$ and $\alpha = 4$ in the path loss function and Nakagami fading with $m = 2$, we can predict the mean and standard deviation of the aggregate interference at node $0$ from $s=0$, i.e., $\Expect{I(t|s=0)}$ and $\operatorname{std}[I(t|s=0)]$, see Fig.~\ref{fig:BrownianMotion2D_alpha4PIwithFading}.
The $\operatorname{std}[I(t|s=0)]$ measures how much the realization of $I(t|s=0)$, i.e., $I(t|t)$, deviates from its predicted mean value, and thus $\operatorname{std}[I(t|s=0)]$ reveals the uncertainty of interference prediction.
In this case, both $\Expect{I(t|s=0)}$ and $\operatorname{std}[I(t|s=0)]$ have closed-form expressions (because~\eqref{eqn:Condition for Some Closed-forms} in Remark~\ref{rek:Condition for Some Closed-forms} is satisfied), which can be derived by calculating $\Expect{I_i(t|s=0)}$ and $\operatorname{std}[I_i(t|s=0)]$, $\forall i$, from~\eqref{eqn:MeanClosedFormAlpha4} and~\eqref{eqn:SecondMomentClosedFormAlpha4} in Appendix~\ref{app:Some Closed Forms for CGPPF}.

From Fig.~\ref{fig:BrownianMotion2D_alpha4PIwithFading} we can see that the mean of the interference prediction does not perform well, and the Coefficient of Variation (CV) $\operatorname{std}[I_i(t|s=0)]/\Expect{I_i(t|s=0)}$, i.e., normalized standard deviation, increases with $t$.
This is due to the fact that the uncertainties ${\bf w}_i(t)$ in velocities $\overline{\bf x}_i(t)$ dominate the behavior of the mobilities of nodes.
In Fig.~\ref{fig:BrownianMotion2DLocations} the node locations are far away from the origin, while prediction tells us $\Expect{{\bf y}_i(t|s=0)} = {\bf y}_i(0) = {\bf 0}$ (see Fig.~\ref{fig:BrownianMotion2DLocations}).
%
%
%

\begin{figure}
\centering
\includegraphics [width=0.72\columnwidth]{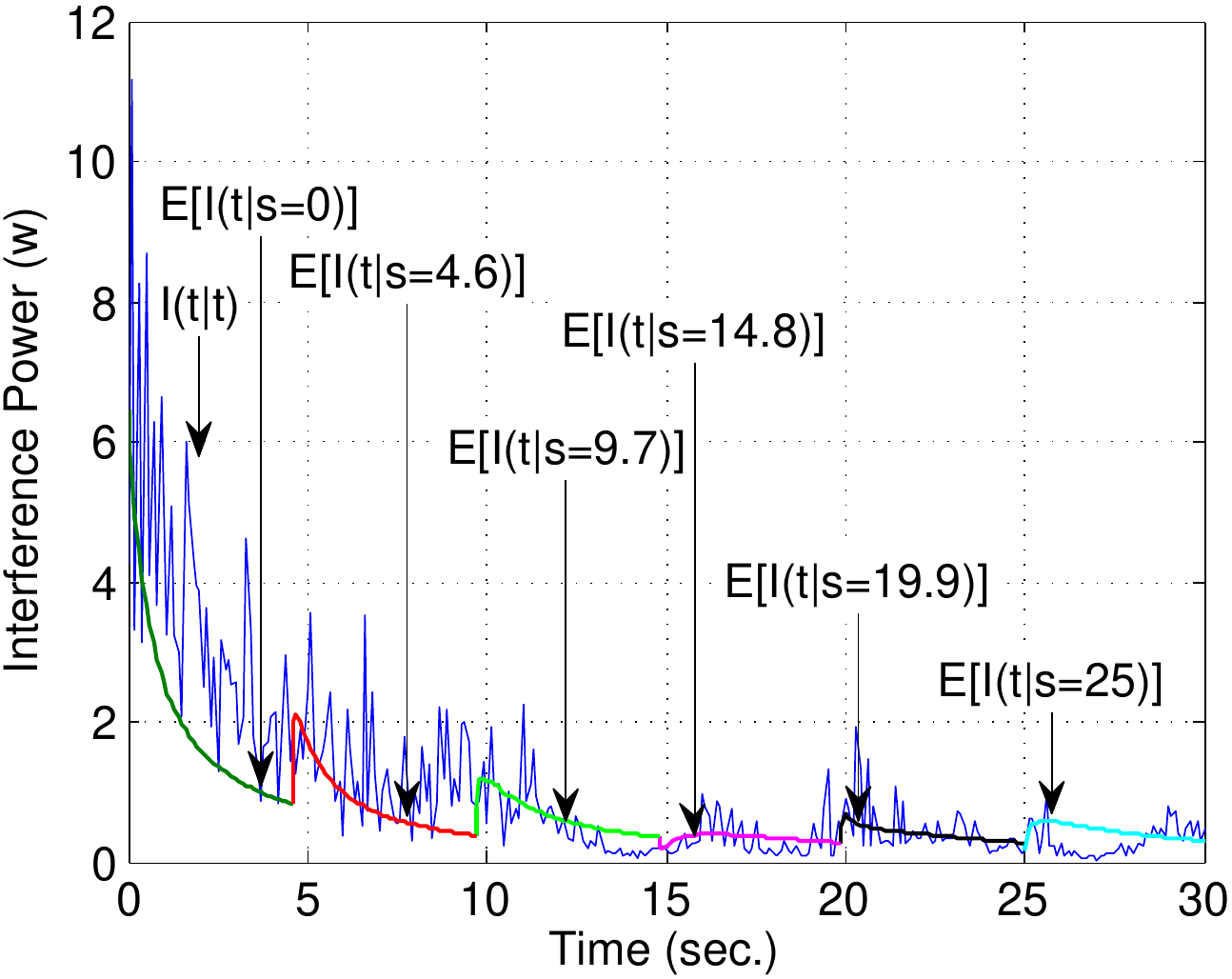}
\caption{Predictions on the mean of aggregate interference based on frequently updated location information at time $s = 0,4.6,9.7,14.8,19.9,25$.}
\label{fig:BrownianMotion2DMultiplePredictions}
\end{figure}

In Fig.~\ref{fig:BrownianMotion2D_alpha4PIwithFading}, $s=0$ means the information on node locations is updated at time $0$.
If we want a better prediction for near future, the information on node locations need to be updated more frequently.
Interference prediction is then done with updated location information until the next update (see Fig.~\ref{fig:BrownianMotion2DMultiplePredictions}).

We end this subsection with a discussion on the Gaussian BPP approximations for these predictions in Fig.~\ref{fig:BrownianMotion2DMultiplePredictions}.
Actually, these predictions can be approximated by a Gaussian BPP at origin.
This is because the condition~\eqref{eqn:NecessaryandSufficientCondtionforGaussianBPPso} in Corollary~\ref{cor:NecessaryandSufficientConditionforGaussianBPPs0} is satisfied.

It should be noted that $\eta_a^{(i)}$ and $\sigma_a^{(i)}$ can be easily calculated to test the condition~\eqref{eqn:NecessaryandSufficientCondtionforGaussianBPPso}:
By \lemref{lem:NodeDistributionunderDynamicReferencePoint}, the mean and variance of the location prediction at $t$ from $s$ for $i$th node are
\begin{align}\label{eqn:MeanandVarianceforBMLocationPrediction}
{\boldsymbol \mu} = {\bf y}_i(s) - {\bf y}_0(s),\quad{\bf \Sigma} = \operatorname{diag}\big\{2(t-s), 2(t-s)\big\},
\end{align}
where ${\bf y}_i(s) - {\bf y}_0(s)$ is finite.
As ${\bf \Sigma}$ is naturally a diagonal matrix (which means the orthogonal transformation ${\bf P}^{T}{\bf\Sigma}^{-1}{\bf P}=\operatorname{diag}\{1/\sigma_1^2,\dotsc,1/\sigma_d^2\}$ is not required), we have
\begin{multline}
\eta_1 = \mu_1 = y_i^{(1)}(s) - y_0^{(1)}(s), \quad\eta_2 = \mu_2 = y_i^{(2)}(s) - y_0^{(2)}(s),\\
\sigma_1 = \sigma_2 = \sqrt{2(t-s)}.~~~~~~~~~~~~~~
\end{multline}
Thus, the condition~\eqref{eqn:NecessaryandSufficientCondtionforGaussianBPPs} in \thmref{thm:NecessaryandSufficientConditionforGaussianBPPs} is satisfied, which implies the predictions made from different $s$ in Fig.~\ref{fig:BrownianMotion2DMultiplePredictions} can be approximated by the corresponding prediction from the Gaussian BPP with $6$ nodes whose location pdf has ${\boldsymbol \mu} = {\bf 0}$ when $t-s$ is large enough.
The approximation has the following close-form expression
\begin{multline}\label{eqn:Mean of Predicted Aggregated Interference Approximated by Gaussian BPP}
\Expect{I(t-s|s)} = 6\Expect{I_i(t-s|s)} \\ \approx \frac {6 \textmd{Ci}\left(\frac {\sqrt{\epsilon}}{2 \sigma^2}\right) \sin\frac {\sqrt{\epsilon}}{2 \sigma^2} + 3\cos\frac{\sqrt{\epsilon}}{2 \sigma^2} \left[\pi - 2\textmd{Si}\left(\frac{\sqrt{\epsilon}}{2 \sigma^2}\right)\right]} {2 \sqrt{\epsilon} \sigma^2},
\end{multline}
where $\Expect{I_i(t-s|s)}$ is derived from~\eqref{eqn:MeanClosedFormAlpha4} in Appendix~\ref{app:Some Closed Forms for CGPPF}, and $\sigma = \sigma_1 = \sigma_2 = \sqrt{2(t-s)}$.
Since $\Expect{I_i(t-s|s)}$ is time-invariant, it is exactly the $\Expect{I(t|s=0)}$.
Therefore, if we translate the time-axis by $s-t$, these prediction $\Expect{I(t|s)}$ ($s = 4.6,9.7,14.8,19.9,25$) will asymptotically converge to $\Expect{I(t|s=0)}$ as $t-s$ increases (see Table~\ref{tab:BPPconvergencebehaviorformeanofpredictedaggregateinterferenceBM}).


\begin{table*}[htb]
\caption{Convergence Behavior for the Mean of the Interference Prediction in Networks with Brownian Motion Mobility models\label{tab:BPPconvergencebehaviorformeanofpredictedaggregateinterferenceBM}}
\centering
\begin{small}
\begin{tabular}{lccccc}
\toprule
~& $t-s = 10$ & $t-s = 50$ & $t-s = 100$ & $t-s = 500$\\
\midrule
$\Expect{I(t|s=0)}$& 0.2201 & 0.0463 & 0.0233 & 0.0047\\
$\Expect{I(t|s=4.6)}$& 0.2115 & 0.0459 & 0.0232 & 0.0047\\
$\Expect{I(t|s=9.7)}$& 0.2032 & 0.0455 & 0.0231 & 0.0047\\
$\Expect{I(t|s=14.8)}$& 0.1772 & 0.0442 & 0.0228 & 0.0047\\
$\Expect{I(t|s=19.9)}$& 0.1768 & 0.0441 & 0.0228 & 0.0047\\
$\Expect{I(t|s=25)}$& 0.1850 & 0.0446 & 0.0229 & 0.0047\\
\bottomrule
\end{tabular}
\end{small}
\end{table*}

\subsection{2D Brownian Motion with Inertia}\label{sec:BrownianMotion2DwithInertia}

The basic Brownian motion in \secref{sec:BrownianMotion2D} is dominated by velocity uncertainty, thus our predictions
cannot be expected to offer great accuracy.
In this section, we focus on the Brownian motion with velocity inertia in 2D space, and it can be employed to describe the motion for vehicle or pedestrian with destination.

\begin{figure}
\centering

\subfigure[]{\includegraphics [width=0.72\columnwidth]{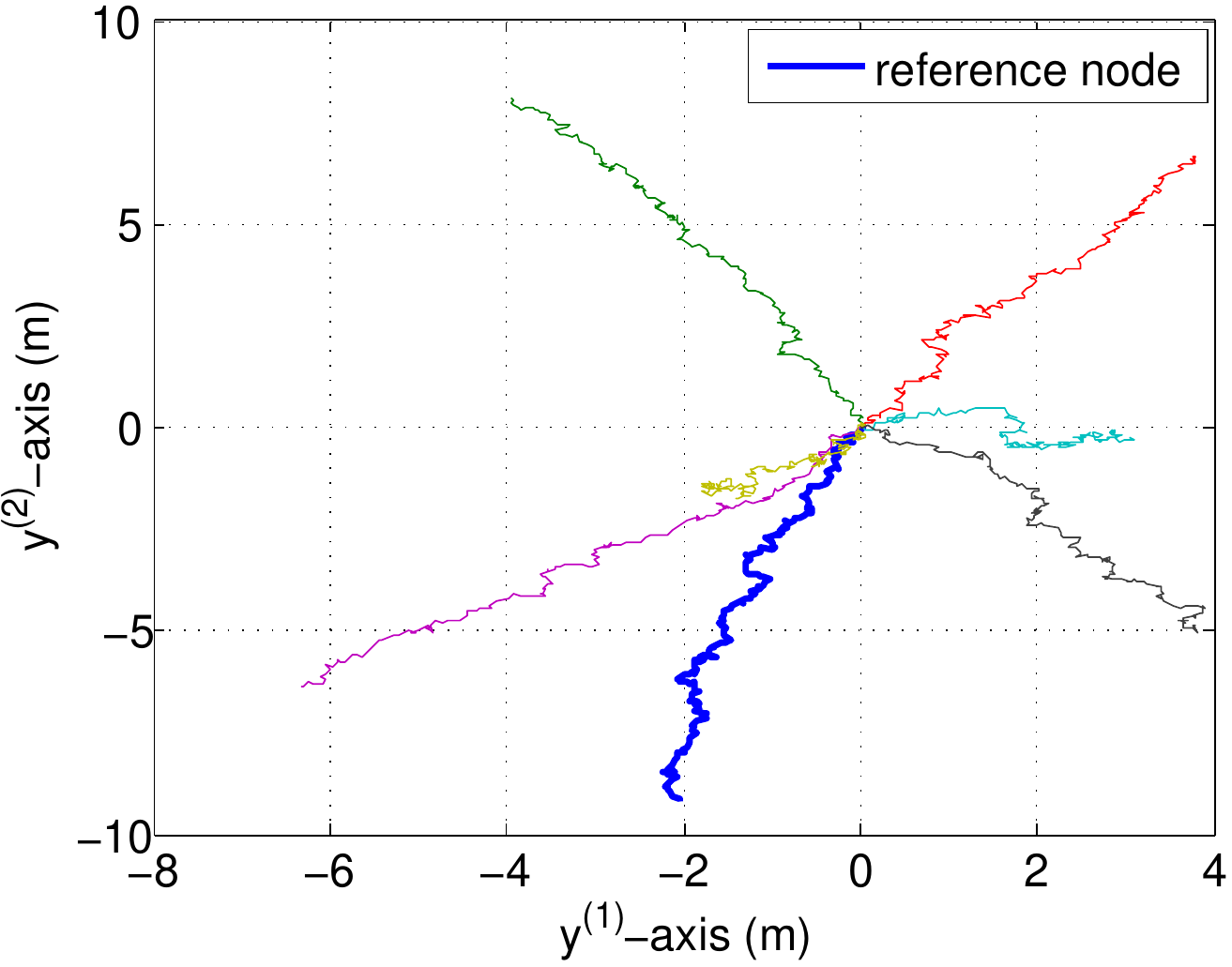}\label{fig:BM2DwithInertiaLocations}}\hfill
\subfigure[]{\includegraphics [width=0.72\columnwidth]{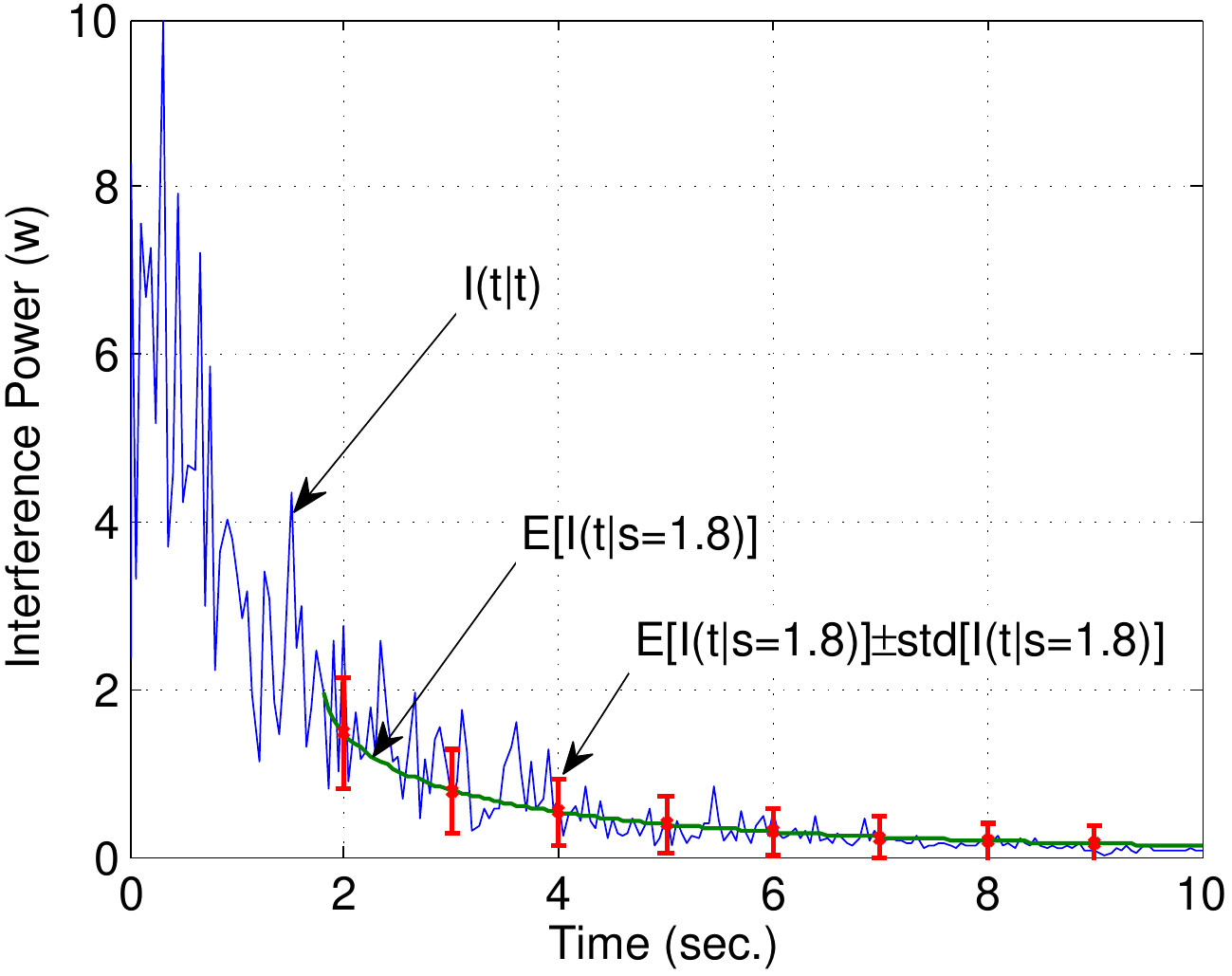}\label{fig:BM2DwithInertiaInterferenceF}}

\caption{(a) Node motions. (b) Aggregate interference at reference node and the mean of the interference prediction.}
\end{figure}

The mobility model has the parameters
\begin{align}\label{eqn:BrownianMotionwithInertiaParameters}
\renewcommand*{\arraystretch}{0.7}
{\bf A}_i = \begin{bmatrix}
0 & 1 & 0 & 0\\
0 & 0 & 0 & 0\\
0 & 0 & 0 & 1\\
0 & 0 & 0 & 0
\end{bmatrix},\quad
{\bf C}_i = \begin{bmatrix}
1 & 0 & 0 & 0\\
0 & 0 & 1 & 0
\end{bmatrix},
\end{align}
where $i=0,1,\dotsc,6$, and with the initial condition
\begin{align}
	{\bf x}_i(0) = \big[0, x_i^{(2)}(0), 0, x_i^{(4)}(0)\big]^T,
\end{align}
where $x_i^{(1)}(0) = 0$ and $x_i^{(3)}(0) = 0$.
Note that the $i$th node location is given by ${\bf y}_i(0) = [x_i^{(1)}(0), x_i^{(3)}(0)]^T$.
The non-zero states, $x_i^{(2)}(0)$ and $x_i^{(4)}(0)$, which represent the $i$th node velocity components (i.e., $\dot{\bf y}_i(0) = [x_i^{(2)}(0), x_i^{(4)}(0)]^T$), are randomly generated in $[-1, 1]$ and assumed to be known.
The ${\bf w}_i$ follows
\begin{align}
	{\bf w}_i(t) = \big[w_i^{(1)}(t), 0\big]^T,
\end{align}
where all $w_i^{(1)}(t)$ are identical GWN with unit power, which implies the node velocities are fluctuating due to these uncertainties.
The parameters of the path loss function and multipath fading are the same as those in \secref{sec:BrownianMotion2D} (i.e., $\epsilon=1$, $\alpha=4$, $m=2$).
A realization for Brownian motion with inertia is shown in Fig.~\ref{fig:BM2DwithInertiaLocations}.
The mean and standard deviation of interference prediction from $s=1.8$ are shown in Fig.~\ref{fig:BM2DwithInertiaInterferenceF}.

From Fig.~\ref{fig:BM2DwithInertiaInterferenceF} we can see that the prediction performs much better than that for Brownian motion without inertia, and the standard deviation is smaller than that in Brownian motion without inertia for the same interference level.
It also should be noted that, similar to the Brownian motion without inertia, the CV becomes larger with increasing $t$.

In terms of the limiting behavior of the prediction, unfortunately, it cannot be approximated by the prediction from a Gaussian BPP in \thmref{thm:NecessaryandSufficientConditionforGaussianBPPs} no matter how large $t$ is.
It can be calculated that
\begin{align}
    \begin{split}
        \eta_1 &= x_i^{(1)}(s) - x_0^{(1)}(s) + \big[x_i^{(2)}(s) - x_0^{(2)}(s)\big] \cdot (t-s)\\
        \eta_2 &= x_i^{(3)}(s) - x_0^{(3)}(s) + \big[x_i^{(4)}(s) - x_0^{(4)}(s)\big] \cdot (t-s)\\
        \sigma_1 &= \sigma_2 = \sqrt{2(t-s)}.
    \end{split}
\end{align}
Because $\big[x_i^{(2)}(s) - x_0^{(2)}(s)\big] \neq \big[x_i^{(4)}(s) - x_0^{(4)}(s)\big]$ at $s=1.8$, condition~\eqref{eqn:NecessaryandSufficientCondtionforGaussianBPPs} in \thmref{thm:NecessaryandSufficientConditionforGaussianBPPs} is not satisfied.
The prediction based on the actual mobility model and that based on BPP approximation are shown in Table~\ref{tab:BPPconvergencebehaviorformeanofpredictedaggregateinterferenceBMI}, and it turns out that there is a big difference between them.

\begin{table*}
\caption{Comparison of the Mean of the Interference Prediction in Networks with BPP Approximation\label{tab:BPPconvergencebehaviorformeanofpredictedaggregateinterferenceBMI}}
\centering
\begin{small}
\begin{tabular}{lcccc}
\toprule
~& $t-s = 10$ & $t-s = 50$ & $t-s = 100$ & $t-s = 500$\\
\midrule
$\Expect{I(t|s=1.8)}$& $2.679\times 10^{-2}$ & $1.837\times 10^{-4}$ & $3.025\times 10^{-5}$ & $8.665\times 10^{-7}$\\
BPP Approximation& 0.2201 & 0.0463 & 0.0233 & 0.0047\\
\bottomrule
\end{tabular}
\end{small}
\end{table*}

\subsection{Uniform Circular Motion in 3D Space}\label{sec:UAVs Target Scanning 3D}

\begin{figure}
\centering
\subfigure[]{\includegraphics [width=0.72\columnwidth, height=0.6\columnwidth]{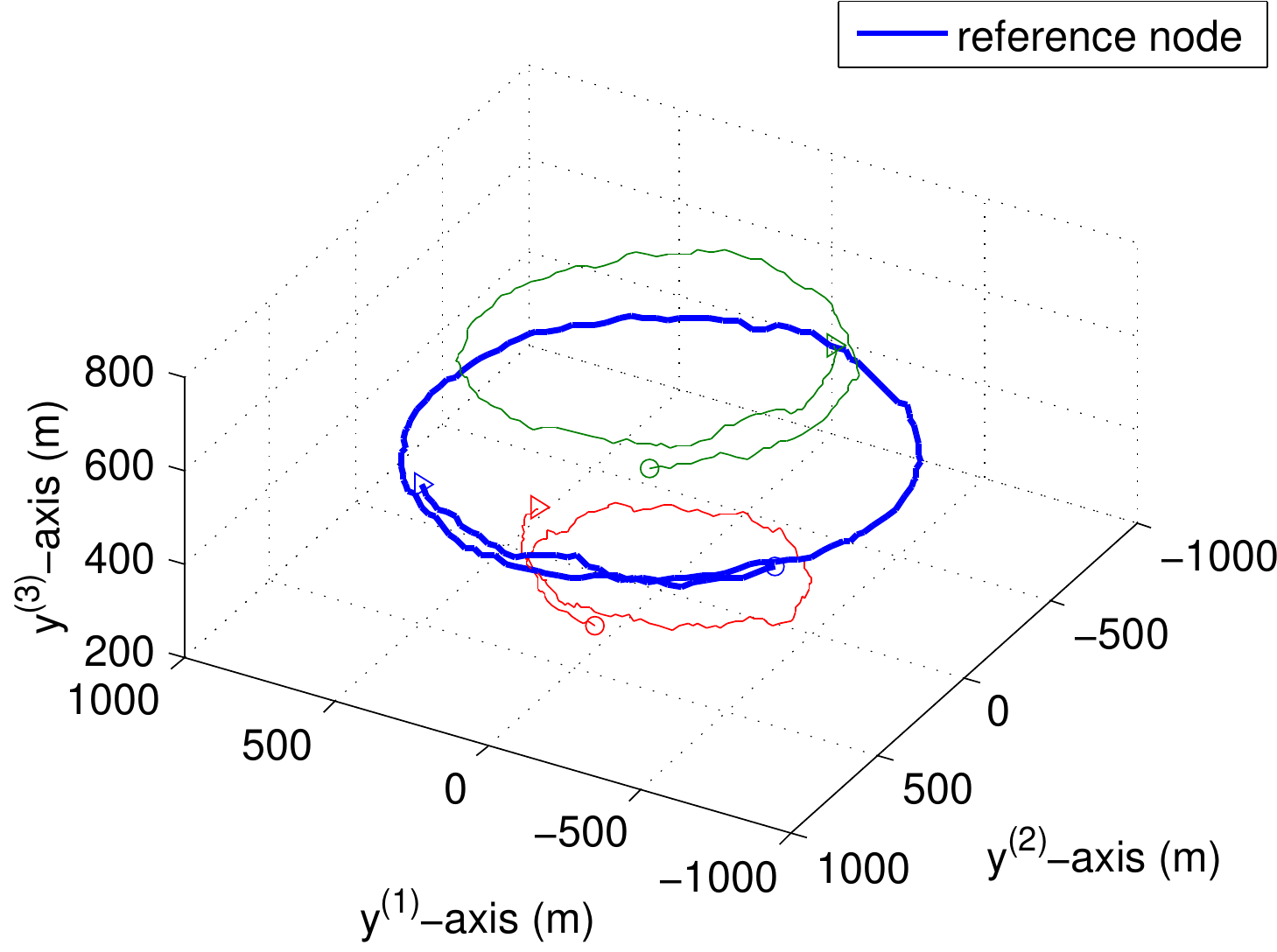}\label{fig:3DUAVsLocations}}\hfill
\subfigure[]{\includegraphics [width=0.72\columnwidth, height=0.6\columnwidth]{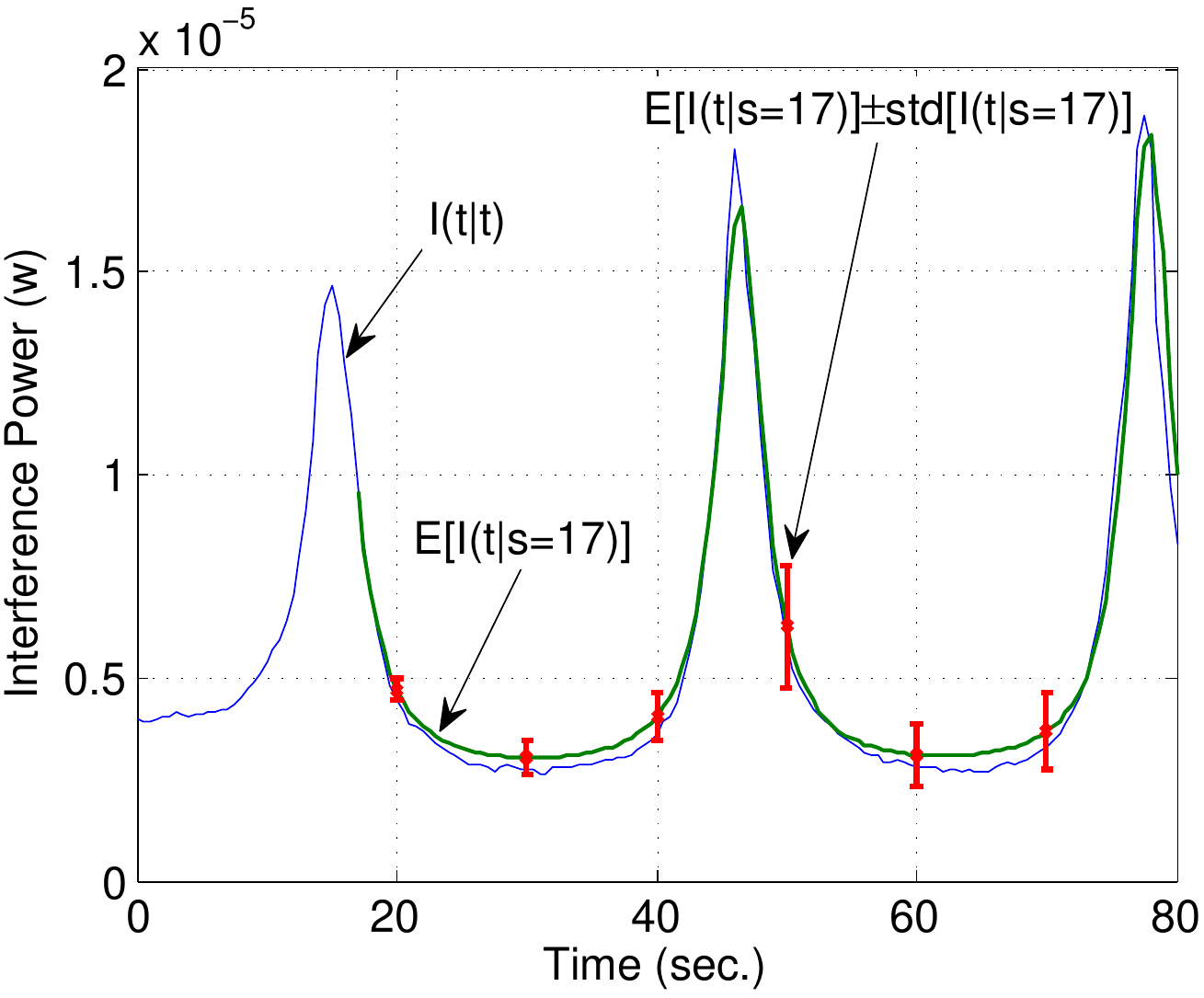}\label{fig:3DUAVsPredictedInterference}}
\caption{(a) UAV locations versus time $t$, starting from the triangular marks and ending at the circular marks. (b) Aggregate interference at reference node and the mean of the interference prediction.}
\end{figure}

In this section, we revisit the UAV target scanning problem discussed in \secref{sec:Uniform Circular Motions}.
In the real world, UAVs that execute the scanning task seldom hover in the same 2D plane (their flight altitudes differ) for collision avoidance.
Thus, UCM should be modeled in 3D rather than 2D.
Assume there is one receiver (reference node, UAV $0$) and two interferers (UAVs $1$ and $2$), whose parameters of mobility models are
\begin{align}\label{eqn:UAVsMobilityModelsParameters}
\renewcommand*{\arraystretch}{0.7}
\begin{split}
{\bf A}_0 &= {\bf A}_2 = \begin{bmatrix}
0 & -0.1 & 0 \\
0.1 & 0 & 0\\
0 & 0 & 0
\end{bmatrix},\quad
{\bf C}_0 = {\bf C}_2 = \begin{bmatrix}
1 & 0 & 0\\
0 & 1 & 0\\
0 & 0 & 1
\end{bmatrix}\\
{\bf A}_1 &= \begin{bmatrix}
0 & 0.1 & 0 \\
-0.1 & 0 & 0\\
0 & 0 & 0
\end{bmatrix},\quad
{\bf C}_1 = \begin{bmatrix}
1 & 0 & 0\\
0 & 1 & 0\\
0 & 0 & 1
\end{bmatrix},
\end{split}
\end{align}
which implies the hover angular velocities for UAVs $0$ and $2$ are both $-0.1$rad/s, and for node $1$ is $0.1$rad/s.
The initial locations ${\bf y}_i(0) = {\bf x}_i(0)$, $i=0,1,2$ are
\begin{align}
\renewcommand*{\arraystretch}{0.7}
{\bf y}_0(0) =
\begin{bmatrix}
500\\
500\\
500
\end{bmatrix},~
{\bf y}_1(0) =
\begin{bmatrix}
-400\\
-300\\
700
\end{bmatrix},~
{\bf y}_0(0) =
\begin{bmatrix}
400\\
0\\
300
\end{bmatrix},
\end{align}
where $y_0^{(3)}(0) = 500$, $y_1^{(3)}(0) = 700$ and $y_2^{(3)}(0) = 300$ are the initial altitudes for UAVs 0, 1, and 2.
The ${\bf w}_i$ follows
\begin{align}
	{\bf w}_i(t) = \big[w_i^{(1)}(t), w_i^{(2)}(t), w_i^{(3)}(t)\big]^T,
\end{align}
where all $w_i^{(j)}(t),j=1,2,3$ are identical GWN with power $\sigma^2 = 100$.
The parameters of path loss function are the same as those in Sections~\ref{sec:BrownianMotion2D} and~\ref{sec:BrownianMotion2DwithInertia} (i.e., $\epsilon=1$, $\alpha=2$).
Because there are few obstacles in airspace, we assume that there is no multipath fading.
The UAVs' motion-trajectories are shown in Fig.~\ref{fig:3DUAVsLocations}.
The aggregate interference and $\Expect[\big]{I(t|s=17)}$ are shown in Fig.~\ref{fig:3DUAVsPredictedInterference}.
We see that the interference prediction is very close to the actual interference.
It is interesting to see that the CV is not always increasing with $t$.
However, if we consider the interference at the same level, the CV does increase with time, e.g., $t=30$ and $t=60$.

\section{Conclusion}

In this paper, the interference prediction problem in MANETs has been investigated.
The GLC mobility model has been proposed to describe or approximate a large class of mobilities in the real world.
The statistics of interference prediction with respect to a dynamic reference point have been analyzed, including mean value and the MGF.
We have defined the CGPPF as a general framework to compute the mean and MGF, and discussed important special cases where closed-form expressions can be obtained.
With expressing the CGPPF in series form, we have analyzed the limiting behavior of the statistics of the interference prediction, and given the necessary and sufficient for when the node locations can be regarded as a Gaussian BPP when analyzing the statistics of interference prediction.

The presented work serves as the first step to develop more comprehensive results in interference predictions.
Even though the CGPPF provides a general framework for calculating the statistics of interference prediction, the closed-form expressions exist only in some special cases.
It is possible to derive closed-form approximations for these statistics using a cumulant-based approach (e.g., similar to~\cite{deLima2012,deLima2013}).
Furthermore, for the limiting behavior of interference prediction with homogenous mobilities, it would be desirable to obtain a more direct link between the mobility model and the condition for the BPP approximations stated in Theorem~\ref{thm:NecessaryandSufficientConditionforGaussianBPPs}.
For example, we observed from our numerical results that if the matrix ${\bf A}$ of the mobility model is Lyapunov-stable~\cite{ChenC1999}, the condition for Gaussian BPP approximation holds, and vise versa.
On the other hand, it would be interesting to extend the results on finite number of nodes to infinite number of nodes and study the condition under which the limiting behavior of the interference prediction converges to that from a PPP.
Another interesting direction for future research is to generalize the assumption for ${\bf w}_i(t)$ beyond Gaussian.
Last but not least, the results on interference prediction obtained in this work can also been used to predict the outage probability at a future time instant.

\appendices

\section{PDF of Node Distribution}\label{app:Proof of PDF of Node Distributions}

A more general proof can be found in~\cite{KalmanR1961}.
For completeness, we provide a proof of \lemref{lem:PDFofNodeDistributions} here:

The solution of~\eqref{eqn:state-equation} is
\begin{align}
{\bf{x}}_i(t) = e^{{\bf A}_i(t-t_0)}{\bf{x}}_i(t_0) + \int_{t_0}^t e^{{\bf A}_i(t-\tau)}{\bf w}_i(\tau)\mathrm{d}\tau\label{eqn:SolutionofStateEquation}
\end{align}
and then the mean of ${\bf{x}}_i(t)$ can be derived as
\begin{align}\label{eqn:MeanofStateDistributions}
\begin{split}
\Expect{{\bf{x}}_i(t)} &= e^{{\bf A}_i(t-t_0)}{\bf{x}}_i(t_0) + \int_{t_0}^t e^{{\bf A}_i(t-\tau)}\Expect{{\bf{w}}_i(\tau)}\,\mathrm{d}\tau\\
&\stackrel {(a)} {=} e^{{\bf A}_i(t-t_0)}{\bf{x}}_i(t_0),
\end{split}
\end{align}
where (a) is established by $\Expect[\big]{{\bf{w}}_i(\tau)} = {\bf 0}$.
Then we can get~\eqref{eqn:MeanofNodeDistributions} from~\eqref{eqn:output-equation}.

The covariance of ${\bf{x}}_i(t)$ is given by
\begin{equation}\label{eqn:ProofofStateVariance1}
\begin{split}
\Cov[\big]{{\bf{x}}_i(t)} &\stackrel{(a)}{=} \Cov[\big]{e^{{\bf A}_i(t-t_0)}{\bf{x}}_i(t_0)} \\&~~~~~~~~~~~+ \Cov[\Big]{\int_{t_0}^t e^{{\bf A}_i(t-\tau)}{\bf w}_i(\tau)\,\mathrm{d}\tau} \\
&= {\bf 0} + \Cov[\Big]{\int_{t_0}^t e^{{\bf A}_i(t-\tau)}{\bf w}_i(\tau)\,\mathrm{d}\tau},
\end{split}
\end{equation}
where (a) follows from the independence of ${\bf{x}}_i(t_0)$ and ${\bf w}_i(\tau)\left(\tau\in[t_0,t]\right)$.
Additionally,
\begin{equation*}
\begin{split}
&\Cov[\Big]{\int_{t_0}^t e^{{\bf A}_i(t-\tau)}{\bf w}_i(\tau)\,\mathrm{d}\tau}\\
&= \Covv[\big]{\int_{t_0}^t e^{{\bf A}_i(t-\tau)}{\bf w}_i(\tau)\,\mathrm{d}\tau, \int_{t_0}^t e^{{\bf A}_i(t-\psi)}{\bf w}_i(\psi)\,\mathrm{d}\psi}\\
&= \int_{t_0}^t \int_{t_0}^t \Covv[\big]{e^{{\bf A}_i(t-\tau)}{\bf w}_i(\tau), e^{{\bf A}_i(t-\psi){\bf w}_i(\psi)}}\,\mathrm{d}\tau\,\mathrm{d}\psi\\
&= \int_{t_0}^t \int_{t_0}^t e^{{\bf A}_i(t-\tau)} \Covv[\big]{{\bf w}_i(\tau), {\bf w}_i(\psi)} e^{{\bf A}_i^T(t-\psi)}\,\mathrm{d}\tau\,\mathrm{d}\psi\\
&\stackrel{(a)}{=} \Theta_{x_i}(t),
\end{split}
\end{equation*}
where (a) follows the properties of GWN.
Then~\eqref{eqn:VarianceofNodeDistributions} can be derived from~\eqref{eqn:output-equation}.

\section{Proof of CGPPF Series Form}\label{app:ProofofGaussianPointProcessFunctionalSeriesForm}

With the orthogonal transform ${\bf z} = {\bf P}^T {\bf y}$ such that ${\bf P}^{T}$\!\!{\boldmath${\mu}$} $ = $ {\boldmath ${\bf \eta}$} $=[\eta_a]_{d\times1}$ and ${\bf P}^{T}{\bf\Sigma}^{-1}{\bf P}=\operatorname{diag}\{1/\sigma_1^2,\dotsc,1/\sigma_d^2\} = {\bf \Lambda}$, equation~\eqref{eqn:GaussianPointProcessFunctional} can be rewritten as
\begin{equation}\label{eqn:Gnuproof}
\begin{split}
G[\nu] &= \int_{\mathbb{R}^d} \nu(\|{\bf y}\|) \frac {1} {(2\pi)^{\frac {d} {2}}|{\bf \Sigma}|^{\frac {1} {2}}} e^{-\frac {1} {2} ({\bf y} - {\boldsymbol \mu})^T {\bf \Sigma}^{-1} ({\bf y} - {\boldsymbol \mu})}\,\mathrm{d}{\bf y}\\
&=\int_{\mathbb{R}^d} \nu(\|{\bf z}\|) \frac {1} {(2\pi)^{\frac {d} {2}}|{\bf \Sigma}|^{\frac {1} {2}}} e^{-\frac {1} {2} ({\bf z} - {\boldsymbol \eta})^T {\bf \Lambda} ({\bf z} - {\boldsymbol \eta})}\,\mathrm{d}{\bf z}\\
&\stackrel {(a)} {=} \frac {1} {(2\pi)^{\frac {d} {2}}|{\bf \Sigma}|^{\frac {1} {2}}} \int_{0}^{\infty} \Big[ \nu(r) \cdot\\ &~~~~~~ e^{-\frac {1} {2} (r{\bf \Phi} - {\boldsymbol \eta})^T {\bf \Lambda} (r{\bf \Phi} - {\boldsymbol \eta})}r^{d-1}\Big] \mathrm{d}r \int_{\bf \Theta} V({\boldsymbol \phi})\,\mathrm{d}{\boldsymbol \phi},
\end{split}
\end{equation}
where (a) follows a $d$-dimensional spherical transform.
${\bf \Phi} = [\Phi_{a}]_{d\times1}$ and $V({\boldsymbol \phi}) \mathrm{d}{\boldsymbol \phi}$ are shown in~\eqref{eqn:Phi} and~\eqref{eqn:ThetaVolumeElement}, respectively.

Labeling
\begin{align}
\overline{\nu}(r, {\bf \Phi}) = \int_{0}^{\infty} \nu(r)\,e^{-\frac {1} {2} (r{\bf \Phi} - {\boldsymbol \eta})^T {\bf \Lambda} (r{\bf \Phi} - {\boldsymbol \eta})} r^{d-1}\,\mathrm{d}r,
\end{align}
we expand it into the Taylor Series
\begin{equation}\label{eqn:nubar}
\begin{split}
\overline{\nu}(r, {\bf \Phi}) &= \sum_{n=0}^{\infty}\frac {(-1)^{n}} {2^n n!} \int_{0}^{\infty} \Big\{ \nu(r) r^{d-1} \cdot\\ &~~~~~~~~~~~~~~~\left[(r{\bf \Phi} - {\boldsymbol \eta})^T {\bf \Lambda} (r{\bf \Phi} - {\boldsymbol \eta})\right]^n \Big\} \mathrm{d}r.
\end{split}
\end{equation}

By employing the multinomial theorem,
\begin{multline}
\left[(r{\bf \Phi} - {\boldsymbol \eta})^T {\bf \Lambda} (r{\bf \Phi} - {\boldsymbol \eta})\right]^n = \sum_{k_1+k_2+k_3 = n}
\Bigg[\binom{n}{k_1,k_2,k_3}\cdot\\
\left(\sum_{a=1}^{d} \frac {\Phi_a^2} {\sigma_a^2} \right)^{\!\!\!k_1} \!\!\!\left(\sum_{b=1}^{d} -\frac {2 \Phi_b \eta_b} {\sigma_b^2}  \right)^{\!\!\!k_2} \!\!\!\left(\sum_{q=1}^{d} \frac {\eta_q^2} {\sigma_q^2} \right)^{\!\!\!k_3} \!\!\!r^{2k_1+k_2}\Bigg],
\end{multline}
where
\begin{align}
\begin{split}
\left(\sum_{a=1}^{d} \frac {\Phi_a^2} {\sigma_a^2} \right)^{\!\!\!k_1} &= \sum_{k_1^{(1)}+\dotsb+k_1^{(d)} = k_1}
\Bigg[\binom{k_1}{k_{1}^{(1)},\dotsc,k_{1}^{(d)}}\cdot\\
&\prod_{a=1}^{d} \left(\frac {\Phi_a} {\sigma_a} \right)^{\!\!\!2k_1}\Bigg],
\end{split}\\
\begin{split}
\left(\sum_{b=1}^{d} -\frac {2 \Phi_b \eta_b} {\sigma_b^2}  \right)^{\!\!\!k_2} &= \sum_{k_2^{(1)}+\dotsb+k_2^{(d)} = k_2}
\Bigg[\binom{k_2}{k_{2}^{(1)},\dotsc,k_{2}^{(d)}}\cdot\\
&\prod_{b=1}^{d} \left(-\frac {2 \Phi_b \eta_b} {\sigma_b^2} \right)^{\!\!\!k_2}\Bigg].
\end{split}
\end{align}
Thus,~\eqref{eqn:Gnuproof} can be written as~\eqref{eqn:Gnu}.

\section{Integrations in CGPPF Series Form}\label{app:IntegrationsinGaussianPointProcessFunctionalSeriesForm}

\subsection{Derivation for $\Psi[\nu]$ in~\eqref{eqn:Psi}}

In order to calculate the interference prediction mean, we set $\nu\left(r\right) = \frac {1} {\epsilon + r^{\alpha}}$, thus $\Psi[\nu]$ is
\begin{equation}
\Psi[\nu] \!=\!
\begin{cases}
\frac {R^{c+1}H_2F_1\left(1, \frac {c+1} {\alpha}, \frac {c+1+\alpha} {\alpha}, -\frac {R^{\alpha}} {\epsilon}\right)} {(1+c)\epsilon} & \epsilon > 0\\
\frac {R^{c-\alpha+1}} {c+\alpha-1} & \epsilon = 0, c-\alpha+1>0,
\end{cases}
\end{equation}
where $c = 2k_1+k_2+d-1$ and $H_2F_1(\cdot)$ is the hypergeometric function~\cite{BatemanH1953}.
When $\epsilon = 0$, the condition $c-\alpha+1>0$ should be satisfied, this is because the singularity at $0$.

In order to calculate the MGF of interference prediction, we set $\nu\left(r\right) = \left[\frac {m} {m - \beta \frac {1} {\epsilon + r^{\alpha}}}\right]^{m}$, thus $\Psi[\nu]$ is
\begin{multline}
\Psi[\nu] = R^{c+1} \left(\epsilon-\beta+R^{\alpha}\right)^{-m} \Bigg[\frac {\epsilon \left(\epsilon-\beta+R^{\alpha}\right) } {(1+c)(\epsilon-\beta)} \cdot \\
H_2F_1\left(1, \frac {1+c+\alpha(1-m)} {\alpha}, \frac {1+c+\alpha} {\alpha}, \frac {R^{\alpha}} {\beta-\epsilon}\right) + \\
\frac {R^{\alpha} \left(1-\frac {R^{\alpha}} {\beta-\epsilon}\right)^m H_2F_1\left(\frac {1+c+\alpha} {\alpha}, m, \frac {1+c+2\alpha} {\alpha}, \frac {R^{\alpha}} {\beta-\epsilon}\right)} {1+c+\alpha}\Bigg],
\end{multline}
where $\epsilon\neq\beta$.

\subsection{Derivation for $\Omega$ in~\eqref{eqn:Omega}}

For $d=1$,
\begin{equation}
\Omega = \frac {(-2)^{k_2} \eta^{k_2 + 2k_3}} {\sigma^{2(k_1+k_2+k_3)}}.
\end{equation}

In order to simplified the discussions for the cas $d = 2$, we define two functions
\begin{multline}
I_{[0, \pi]}(m,n) = \\
\begin{cases}
\sum_{l=0}^{\frac {m} {2}}
\binom{m/2}{l}
(-1)^{l} \frac {\sqrt{\pi}\Gamma(\frac {2l+n+1} {2})} {\Gamma(\frac {2l+n+2} {2})} & m \text{ is even}, m > 0, n \geq 0\\
0 & m \text{ is odd}, m > 0, n \geq 0\\
\frac {\sqrt{\pi}\Gamma(\frac {n+1} {2})} {\Gamma(\frac {n+2} {2})} & m=0, n \geq 0
\end{cases}
\end{multline}
and
\begin{multline}
I_{[0, 2\pi]}(m,n) =\\
\begin{cases}
\Upsilon (m,n) & n \text{ is even}, m \geq 0, n > 0\\
0 & n \text{ is odd}, m \geq 0, n > 0\\
\frac {\sqrt{\pi}\Gamma(\frac {n+1} {2})} {\Gamma(\frac {n+2} {2})} & m \geq 0, n = 0
\end{cases},~~~~~
\end{multline}
where
\begin{align*}
\Upsilon (m,n) = \sum_{l=0}^{\frac {n} {2}}
\binom{n/2}{l}(-1)^{l} \frac {\left[1+(-1)^{2l+m}\right]\sqrt{\pi}\Gamma(\frac {2l+m+1} {2})} {\Gamma(\frac {2l+m+2} {2})}.
\end{align*}

Therefore, with $d=2$, we have
\begin{align}
\Omega = \left[\sum_{i=1}^{2} \left(\frac {\eta_{i}} {\sigma_i}\right)^{2} \right]^{k_3}\!\!\!\!\!\!\sum_{k_{1}^{(1)}+k_{1}^{(2)}=k_1 \atop k_{2}^{(1)}+k_{2}^{(2)}=k_2}
\binom{k_1}{k_{1}^{(1)},k_{1}^{(2)}}
\binom{k_2}{k_{2}^{(1)},k_{2}^{(2)}}
\,\Xi,
\end{align}
where
\begin{multline}
\Xi = \bigg[\prod_{1\leq i\leq 2 \atop 1\leq j\leq 2} \left(\frac {1} {\sigma_i}\right)^{2k_{1}^{(i)}} \left(-2\frac{\eta_{j}}{\sigma_j^2}\right)^{k_{2}^{(j)}}\bigg]\cdot\\
\Bigg[I_{[0, 2\pi]}\left(2k_{1}^{(1)}+k_{2}^{(1)},0\right) I_{[0, 2\pi]}\left(2k_{1}^{(1)},k_{2}^{(2)}\right)\cdot\\ I_{[0, 2\pi]}\left(k_{2}^{(1)}, 2k_{1}^{(2)}\right) I_{[0, 2\pi]}\left(0,2k_{1}^{(2)}+k_{2}^{(2)}\right)\Bigg].
\end{multline}

\section{Some Closed Forms for CGPPF}
\label{app:Some Closed Forms for CGPPF}

If condition~\eqref{eqn:Condition for Some Closed-forms} is satisfied, some closed-form expressions for $\Expect{I_i(t|s)}$ and $\Expect{I_i^2(t|s)}$ can be derived.
Please refer to~\eqref{eqn:PathLossFunction} for the definition of $\alpha$ and $\epsilon$.

Firstly, if $\epsilon = 0$, which implies a singular path loss, we can derive the closed-form expressions for first and second order statistics as
\begin{align}
	\Expect[\big]{I_i(t|s)} &= \int_{0}^{\infty} \frac {v_d r^{d-1}} {(2\pi)^{\frac {d} {2}} \sigma^{d} r^{\alpha}} e^{-\frac {r^2} {2\sigma^2}} \mathrm{d}r = \frac {v_d \Gamma\big[\frac {d - \alpha} {2}\big]} {2^{\frac {\alpha} {2}+1} {\pi}^{\frac {d} {2}} \sigma^{\alpha}}\label{eqn:MeanClosedFormSingularPathLoss}\\
\begin{split}
	\Expect[\big]{I_i^2(t|s)} &= \frac {m+1}{m} \int_{0}^{\infty} \frac {v_d r^{d-1}} {(2\pi)^{\frac {d} {2}} \sigma^{d} r^{2\alpha}} e^{-\frac {r^2} {2\sigma^2}} \mathrm{d}r \\&= \frac {(m+1)v_d \Gamma\big[\frac {d - 2\alpha} {2}\big]} {m 2^{\alpha+1} {\pi}^{\frac {d} {2}} \sigma^{2\alpha}},\label{eqn:SecondMomentClosedFormSingularPathLoss}
\end{split}
\end{align}
when gamma function $\Gamma(\cdot)$ has finite value.
%
%
$v_d$ is the volume of the $d$-dimensional ball of radius 1, i.e.,
\begin{align}\label{eqn:VolumeofdDimensionalBall}
v_d = \frac {2\pi^{d/2}} {\Gamma(d/2)}.
\end{align}

\newcommand{\sbinom}[2]{\genfrac{}{}{0pt}{}{#1}{#2}}

However, if $\epsilon > 0$, closed-forms are difficult to derive for general $\alpha > 0$.
We just give the results for $\alpha = 2$ and $\alpha = 4$.
When $\alpha = 2$
\begin{align}
&\Expect[\big]{I_i(t|s)} = \frac {v_d e^{\frac {\epsilon} {2\sigma^2}} \epsilon^{\frac {\alpha - 2} {2}} \Gamma(d/2) \Gamma(\frac {2-d} {2}, \frac {\epsilon} {2\sigma^2})} {2(2\pi)^{\frac {d} {2}}|{\bf \Sigma}|^{\frac {1} {2}}}\label{eqn:MeanClosedFormAlpha2}\\
\begin{split}
&\Expect[\big]{I_i^2(t|s)} = \frac {(m+1) v_d (d-4) \Gamma\left(\frac {d-4} {2}\right) } {16m(2\pi)^{\frac {d} {2}}|{\bf \Sigma}|^{\frac {1} {2}} \epsilon^2 \sigma^4} \cdot \\
&\left\{2\sigma^2 e^{\frac {\epsilon} {2\sigma^2}} \epsilon^{\frac {d} {2}} \left[\epsilon + \sigma^2(d-2)\right] \Gamma\left(\frac {4 - d} {2}, \frac {\epsilon} {2\sigma^2}\right) - 2^{\frac {d} {2}} \sigma^d \epsilon^2\right\}.\label{eqn:SecondMomentClosedFormAlpha2}
\end{split}
\end{align}
For $\alpha = 4$ and $d = 2$,
\begin{align}
	\Expect[\big]{I_i(t|s)} &= \frac {2 \textmd{Ci}\left(\frac {\sqrt{\epsilon}}{2 \sigma^2}\right) \sin\frac {\sqrt{\epsilon}}{2 \sigma^2} + \cos\frac{\sqrt{\epsilon}}{2 \sigma^2} \left[\pi - 2\textmd{Si}\left(\frac{\sqrt{\epsilon}}{2 \sigma^2}\right)\right]} {4 \sqrt{\epsilon} \sigma^2}\label{eqn:MeanClosedFormAlpha4}\\
\Expect[\big]{I_i^2(t|s)} &= \frac {(m+1) G_{1,3}^{1,3}\left(\sbinom{1/2}{0,1/2,3/2}\left|
\frac {\epsilon} {16\sigma^4}
\right.\right)} {4m \epsilon^{\frac{3}{2}} \sqrt{\pi} \sigma^2},\label{eqn:SecondMomentClosedFormAlpha4}
\end{align}
where $\textmd{Si}(\cdot)$ and $\textmd{Ci}(\cdot)$ are sine/cosine integral function with the form
\begin{equation*}
\textmd{Si}(z) = \int_{0}^{z} \frac {\sin x} {x}\,\mathrm{d}x, \text{\quad and\quad}
\textmd{Ci}(z) = -\int_{z}^{\infty} \frac {\cos x} {x}\,\mathrm{d}x,
\end{equation*}
and $G_{m,n}^{p,q}\left( \sbinom{a_1,\dotsc,a_p}{b_1,\dotsc,b_q}\left|z
\right.\right)$ is the Meijer-G function~\cite{BatemanH1953}.

\bibliographystyle{IEEEtran}

\bibliography{InterferencePrediction}

%




\end{document}